\documentstyle[12pt,epsf]{article}

\newcommand{\tr}{\mbox{Tr}}

\let\a=\alpha \let\be=\beta  
   
  \let\la=\lambda 
    \let\s=\sigma

\def\0{\over } 
\def\1{\vec }     
\def\2{{1\over2}} 
\def\4{{1\over4}}            
\def\5{\bar }  
\def\6{\partial } 
\def\7#1{{#1}\llap{/}}                         
\def\8#1{{\textstyle{#1}}}
\def\9#1{{\bf {#1}}}                           
 
\def\llp{\hbox to 0pt{\hss/\hskip1.5pt}}
\def\llo{\hbox to 0.2pt{\hss /}} \def\llq{\hbox to 0pt{\hss/\hskip0.5pt}}
\def\so{\supset\hbox to 0pt{\hss $\displaystyle -$\hskip1pt}}

\def\<{\langle } \def\>{\rangle }

   \let\hc=\dagger

\let\nn=\nonumber  
\def\bea{\begin{eqnarray}} \def\eea{\end{eqnarray}} 
\def\beann{\begin{eqnarray*}} \def\eeann{\end{eqnarray*}} 
\def\beq{\begin{equation}} \def\eeq{\end{equation}}  
\begin{document} 
\setlength{\baselineskip}{18pt}                                     
\thispagestyle{empty}
\begin{flushright}
{\tt OUTP-95-40P\\ DESY-96-017 \\ February 1996}
\end{flushright}
\vspace{5mm}
\begin{center}
{\Large \bf
 On the Mass Spectrum of the SU(2) Higgs Model in 
 2+1 Dimensions}\\ \vspace{15mm}
{\large O.~Philipsen$^{1}$, M.~Teper$^{1}$
 and H.~Wittig$^{2}$}\\ \vspace{10mm}
{\it $^{1}$Theoretical Physics, University of Oxford \\1 Keble Road,
            Oxford OX1 3NP, U.K.\\
\vspace{5mm}
$^{2}$ DESY-IfH Zeuthen,\\ Platanenallee 6, D-15738 Zeuthen, Germany}

\end{center}
\vspace{2cm}

\begin{abstract}
\thispagestyle{empty}
\noindent
We calculate the masses of the low-lying states with quantum numbers
$J^{PC}=0^{++},1^{--}$ in the Higgs and confinement regions of the
three-dimensional SU(2) Higgs model, which plays an important r\^ole in
the description of the thermodynamic properties of the standard model
at finite temperatures.  We extract the masses from correlation
functions of gauge-invariant operators which are calculated by means
of a lattice Monte Carlo simulation. The projection properties of our
lattice operators onto the lowest states are greatly improved by the
use of smearing techniques.  We also consider cross correlations
between various operators with the same quantum numbers. From these
the mass eigenstates are determined by means of a variational
calculation.  In the symmetric phase, we find that some of the ground 
state masses
are about 30\% lighter than those reported from previous
simulations. We also obtain the masses of the first few excited states
in the symmetric phase.  Remarkable among these is the occurrence of a
$0^{++}$ state composed almost entirely of gauge degrees of
freedom. The mass of this state, as well as that of its first
excitations, is nearly identical to the corresponding glueball states
in three-dimensional SU(2) pure gauge theory, indicating an
approximate decoupling of the pure gauge sector from the Higgs sector
of the model. We perform a detailed study of finite size effects and
extrapolate the lattice mass spectrum to the continuum.
\end{abstract} 
\setcounter{page}{0}

\newpage

\section{Introduction}
The study of three-dimensional field theories has attracted a lot of 
attention over the past few years. While some models are investigated
for field theoretic reasons or because they are more easily 
accessible than their four-dimensional homologues, others have an 
immediate physical meaning in the context of 
four-dimensional field theory at finite temperature.
It has been known for a long time that for temperatures much higher than
any mass scale of a given theory its non-static Matsubara modes may be
integrated out perturbatively to yield a three-dimensional effective
theory for the zero modes \cite{appel}. 
This effective theory describes the static long-range physics 
of the underlying four-dimensional finite-temperature theory,
and moreover contains all the infrared divergences and
non-perturbative phenomena that spoil a purely perturbative treatment
of the latter.

In particular, the three-dimensional SU(2) Higgs model represents
an effective high-temperature theory for the standard electroweak
model, after neglecting the U(1) sector and fermions in a first
approximation. 
Since it was realised that the baryon asymmetry of
the universe could have been generated during a first-order
electroweak phase transition \cite{krs85}, a lot of effort has been
spent to determine the order and dynamics of this transition in detail.
The perturbative procedure of dimensionally reducing
the four-dimensional SU(2) Higgs model at finite temperature 
to a three-dimensional effective theory has been carried
out in great detail in ref.~\cite{fkrs94}, and the corresponding
relations between the three-dimensional and four-dimensional
parameters and temperature may be found there.  Bearing these
relations in mind, we shall stay entirely within the framework of the 2+1
dimensional SU(2) Higgs model in this paper.

There are already several analytical \cite{bp94}-\cite{wet93} and
numerical \cite{kaj93}-\cite{kaj95} studies of the three-dimensional
SU(2) Higgs model.  While the main motivation for these studies was the
phase transition itself, there also emerged the problem of
understanding the structure of the symmetric phase. 
Due to infrared divergences in vector loops, 
straightforward perturbation theory breaks down in the
symmetric phase, and until recently not much was known about the
particle spectrum and the effective interactions in this parameter
region of the theory. In a recent analytic calculation, the loop
expansion was reorganised by resumming masses and vertices, which led
to a set of gauge independent gap equations for the vector boson and
the Higgs masses, defined on their respective mass shells.  The
solutions of these equations predict a non-vanishing vector boson mass
and scalar vacuum expectation value in the ``symmetric phase", which
thus would be interpreted as another Higgs phase, just with
different parameters. On the other hand, lattice simulations in four
dimensions at finite temperature \cite{mon95} as well as in three
dimensions \cite{ilg95,kaj95} have reported vector boson masses about 
four times larger than predicted by the gap equations.\footnote{
For a detailed comparison of the lattice and analytic 
approaches in three and four dimensions in the context of
the electroweak phase transition, see \cite{jan95}.}
The picture conjectured from the lattice results is one of a symmetric
phase with confining behaviour (in the sense of QCD) 
and a particle spectrum consisting of
bound states.  Similar conclusions may be drawn from analytic
investigations of truncated renormalisation group equations which
indicate strong coupling effects in the symmetric phase \cite{wet93}.
The picture of a QCD-like symmetric electroweak phase has also been
employed for a model calculation of bound state masses \cite{do95}.
One possible explanation of the large discrepancy between the two
approaches is a breakdown of the resummed loop expansion of
\cite{bp94} in higher orders. 
In principle, however, it is also
conceivable that an extremely low-lying state might not have been visible on
the lattice sizes that have been investigated. Moreover, all simulations have
emphasised the difficulty of measuring correlation functions in the
symmetric phase due to the extremely low signal-to-noise ratio.

The purpose of the present paper is to shed more light on the
situation in the symmetric phase by employing
new techniques which allow a more reliable lattice calculation
of the mass spectrum. The masses are extracted from
correlation functions of gauge invariant operators.  In order to
improve the sensitivity to low-lying or bound states we construct a
large set of non-local operators by employing a ``blocking" technique
similar to the one which has proved to be useful in pure gauge
calculations \cite{tep87,tep92}. As we shall see, this procedure turns
out to be very effective in enhancing the projection of our operators
onto the lowest states.  Moreover, it reduces the statistical errors
significantly, yielding more accurate results for the masses.  We also
measure cross correlations between different operators.
Diagonalisation of the corresponding correlation matrix then unmixes
the superposition of the ground state and the excited states.
This procedure further improves
the signal for the lowest states and,
more importantly, enables us to estimate
the masses of the first few excited states.  It also allows us to extract
information about the overlap any individual operator has with a given
state, and hence the coupling between the different states.  We then
perform a study of finite-size effects and an extrapolation to the
continuum limit for two points in parameter space, one located in the
symmetric and one in the Higgs phase.
 
The paper is organised as follows. In section\,\ref{sec:action} the
lattice action and the basic operators used in mass calculations are
discussed. The details and more technical aspects of our simulation
are described in section\,\ref{sec:simulation}. In
section\,\ref{sec:results} we present our results, analysing in detail
the reliability of our mass estimates of the ground state, and
including the extrapolation to the continuum limit. Finally,
section\,\ref{sec:conclusions} contains our conclusions.

\section{Action and basic operators}
\label{sec:action}

The action of the SU(2) Higgs model in 2+1 dimensions and its general
properties in the continuum and on the lattice have been discussed
previously in the literature \cite{bp94}-\cite{kaj95}.  In order to
fix the notation and to give all equations used in this paper we list
some of these general aspects here.

The continuum action is given by 
\beq\label{l3d}
S = \int d^3x \; \tr \left[{1\over 2}W_{\mu\nu}W_{\mu\nu} +
(D_{\mu}\varphi)^\hc D_{\mu}\varphi + \mu_3^2 \varphi^\hc \varphi
+ 2 \lambda_3 (\varphi^\hc \varphi)^2 \right] \, ,
\eeq
where all fields are in a $2\times2$ matrix notation,
\beq
\varphi = {1\over 2} (\s + i \vec{\pi}\cdot \vec{\tau}) \, ,\quad
D_{\mu}\varphi = (\partial_{\mu} - i g_3 W_{\mu})\varphi\, ,\quad
W_{\mu} = {1\over 2}\vec{\tau}\cdot \vec{W_{\mu}}\ .
\eeq
The gauge coupling $g_3$ and the scalar coupling $\lambda_3$ have mass
dimension 1/2 and 1, respectively.  The action (\ref{l3d}) can be
parametrised by two dimensionless parameters, which may be chosen to
be $\lambda_3/g_3^2$ and $\mu_3^2/g_3^4$. Fixing these parameters
determines the physical properties of the theory.  The corresponding
lattice action may be defined as
\bea \label{actlat}   
S[U,\phi]&=&\be_G\sum_p\left(1-\frac{1}{2}\tr U_p\right)
+\sum_x\Bigg\{-\be_H\sum_{\mu=1}^3\frac{1}{2}
\tr\Big(\phi^{\dagger}(x)U_\mu(x)
\phi(x+\hat{\mu})\Big)  \nn\\
&&+\frac{1}{2}\tr\Big(\phi^{\dagger}(x)\phi(x)\Big)
+\be_R\left[\frac{1}{2}\tr\Big(\phi^{\dagger}(x)\phi(x)\Big)-1\right]^2
\Bigg\}.
\eea

Due to the super-renormalisability of the theory (\ref{l3d}), the only
parameter receiving ultraviolet renormalisation is the scalar mass
parameter $\mu_3^2/g_3^4$, whose corrections have been determined
at the two-loop level \cite{fkrs94} in perturbation theory
using the $\overline{MS}$ scheme.  The corresponding two-loop
calculation in lattice perturbation theory was carried out in
\cite{lai95}.  Requiring that the renormalised mass parameters be the
same in both regularisation schemes, a relation between the
parameters labelling the continuum and lattice theories has been
established \cite{lai95},
\bea \label{lcp1}
\be_G&=&\frac{4}{ag_3^2},\\
\be_R&=&\frac{\lambda_3}{g_3^2} \frac{\be_H^2}{\be_G},\\
\frac{\mu_3^2}{g_3^4}&=&\frac{\be_G^2}{8}\left(\frac{1}{\be_H}-3
-\frac{2\be_H}{\be_G}\frac{\lambda_3}{g_3^2}
\right)+\frac{3\Sigma\be_G}{32\pi}\left(1+
4\frac{\lambda_3}{g_3^2}\right)\nn \\
& &+\frac{1}{16\pi^2}\left[\left(\frac{51}{16}+9\frac{\lambda_3}{g_3^2}
-12\left(\frac{\lambda_3}{g_3^2}\right)^2\right)
\left(\ln\frac{3\be_G}{2}+\zeta\right)+5.0+5.2\frac{\lambda_3}{g_3^2}\right] 
\; ,\label{lcp3}
\eea
with the numerical constants $\Sigma=3.17591$ and $\zeta=0.09$.

A Monte Carlo simulation of any quantity in the theory (\ref{actlat})
is carried out for a given set of bare parameters $\be_G,\be_H,\be_R$.
In order to establish contact with the desired 
continuum physics one first has to
perform an infinite volume limit, i.e., do simulations on lattices
much larger than the largest correlation length of the theory such
that the results do not show any dependence on the lattice size.
Secondly, one has to perform a continuum limit $a\rightarrow 0$, i.e.,
to simulate at different values of $\be_G$ and extrapolate to
$\be_G\rightarrow \infty$. This limit has to be taken in such a way
that the renormalised quantities parametrizing the theory remain
constant.  The corresponding `lines of constant physics' in the space
of the lattice parameters are given by equations
(\ref{lcp1})-(\ref{lcp3}).

The actions (\ref{l3d}) and (\ref{actlat}) have an $\rm
SU(2)_{local}\times SU(2)_{global}$ symmetry.  Physical states are
described by gauge invariant operators.  After decomposing $\phi(x)$
as
\beq
\phi(x)=\rho(x)\a(x),\quad 
\rho^2(x)=\frac{1}{2}\tr\left(\phi^{\dagger}(x)\phi(x)\right)\;,\quad
\rho(x)\ge 0,\quad \a(x) \in SU(2)\;,
\eeq
one may define the gauge-invariant composite field
\beq
V_{\mu}(x)=\a^{\dagger}(x)U_\mu(x)\a(x+\hat{\mu})\;.
\eeq
%

While $\rho(x)$ and $V_{\mu}(x)$ are invariant under local transformations,
they transform under the diagonal global $\rm SU(2)_{diag}$ subgroup,
customarily termed weak isospin, as
\beq
\rho'(x)=\rho(x),\quad V'_{\mu}(x)=\Lambda V_{\mu}(x)\Lambda^{-1},
 \quad \Lambda \in
\rm SU(2)_{diag}\;,
\eeq 
i.e., the lowest excitation of $\rho(x)$ describes the isoscalar Higgs
boson while the matrix-valued $V_\mu(x)$ transforms as an isovector. 
A single field representing the
spin-one, isospin-one $W$ boson may be obtained from the composite link variable
by taking the trace with an insertion of a Pauli matrix $\tau^a$.
Taking the trace without $\tau^a$-insertion produces another spin zero
isoscalar operator.  A third $0^{++}$ isoscalar operator is given by
the plaquette.  Thus we consider the following set of basic operators
for the description of the low-lying states,
\bea \label{ops}
0^{++}:\;R(x) & \equiv &\frac{1}{2}\tr\left(\phi^{\dagger}(x)\phi(x)\right),
 \nonumber \\ 
0^{++}:\;L(x) & \equiv & \sum_{\mu=1}^2 
\frac{1}{2}\tr\left(\a^{\dagger}(x)U_\mu(x)\a(x+\hat{\mu})\right),
 \nonumber \\ 
0^{++}:\;P(x )& \equiv & U_1(x)U_2(x+\hat{1})U^{\dagger}_{1}(x+\hat{2})
U^{\dagger}_{2}(x), \nn \\
1^{--}:\;V_{\mu}^a(x) & \equiv &
\frac{1}{2}\tr\left(\tau^a\a^{\dagger}(x)U_\mu(x)\a(x+\hat{\mu})\right).
\eea
The plaquette $P$ is particularly interesting because it
consists of gauge degrees of freedom only, and in the 
three-dimensional pure gauge
theory it is the simplest operator one can use to describe 
the $0^{++}$ glueball \cite{tep92}.  In
the theory with scalars one expects a mixing of this operator with the
other $0^{++}$ operators due to the coupling of gauge and scalar 
degrees of freedom.

The phase structure of the three-dimensional model with lattice action
(\ref{actlat}) has not been fully mapped out by numerical simulations
as yet. However, from analyticity considerations \cite{frad79} and numerical
studies of the four-dimensional model \cite{kue84}
one expects the following qualitative picture: the three-dimensional
parameter space spanned by $\be_G,\be_H,\be_R$ is divided into Higgs
and confinement-like (or symmetric) regions by a surface of 
first-order phase transitions which is crossed by changing $\be_H$ for fixed
$\be_G,\be_R$. At sufficiently large values of 
$\be_R$ and small values of $\be_G$
this surface is expected to terminate so that the two regions are
analytically connected. In this region, there is no phase transition
but just a crossover as $\be_H$ is varied.  Numerically, however, this
region in the phase diagram has so far not been accessed in the
three-dimensional theory. The continuum limit is represented by a
single point in the phase diagram, $\be_G\rightarrow \infty,
\be_R\rightarrow 0, \be_H\rightarrow 1/3$. In order to describe
different continuum theories, the continuum limit has to be taken
along different paths in the parameter space, as specified by equations
(\ref{lcp1})-(\ref{lcp3}).

The term confinement-like is chosen to distinguish the behaviour of
the theory in this region of parameter space from the confinement
realised in the three-dimensional pure gauge theory.  There the
potential between static charges rises linearly with distance, without
any bound.  In the Higgs model one expects a flattening of the
potential at some large distance, due to pair creation of scalars
breaking the string between the static charges,
just as fermions break the string in QCD. From the analytic
connectedness of the Higgs and the confinement regions it follows that
for every state in the Higgs region, there is a corresponding one in
the confinement region. In particular, the same operators (\ref{ops})
may be used to describe physical states in both regimes.  The global
isospin symmetry is realised in the Higgs as well in the confinement
region, so one expects low-lying states with the same quantum numbers
in both regions.

\section{The Simulation}
\label{sec:simulation}

The purpose of this paper is a closer investigation of the mass
spectrum in the confinement-like phase, by calculating correlations of
operators of the type in eq.\,(\ref{ops}).

In this section, we describe the details of our calculation, including
the simulation algorithm, the construction of ``blocked" or ``smeared"
operators for the Higgs and vector bosons and the way in which these are
used to compute matrix correlators. We conclude this section with
details about the statistical analysis and the fitting procedure
employed to obtain our final mass estimates.

\subsection{Simulation algorithm and parameters}
\label{sec:algol}

Our Monte Carlo simulation is performed using the lattice action in
eq.\,(\ref{actlat}), containing the bare parameters
$\beta_G,\,\beta_H$ and $\beta_R$.

For the update of the gauge variables we use a combination of the
standard heatbath and over-relaxation algorithms for SU(2)
\cite{fabhaan,kenpen}. The scalar degrees of freedom are updated
using the algorithm proposed in \cite{bunk_lat94}, which uses the four
real components of the scalar field $\phi(x)$. Thus, no separate
updates of the radial and angular parts $\rho(x)$,$\alpha(x)$ are
required, leading to a simple implementation of the algorithm. As
explained in \cite{bunk_lat94}, over-relaxation (reflection) steps in
the update of the scalar field can be easily incorporated, provided
the Higgs self-coupling $\beta_R$ is not too large, which would lead
to a poor acceptance rate. In our simulation, where $\beta_R =
O(10^{-4})$, we achieved acceptance rates of well over 90\%.  Higher
values of $\beta_R$ could be simulated, for instance, by using the
reflection algorithm described in \cite{fodjan}.

In our simulation, a ``compound" sweep consists of a combination of
heatbath (HB) and several reflection (REF) updates of the gauge and scalar
fields,
\beq
 1\,{\rm HB}\big\{U\big\} + 1\,{\rm HB}\big\{\phi\big\} + n_{OR}\Big\{
{\rm REF}\big\{U\big\} + n_{\rm ref} {\rm REF}\big\{\phi\big\} \Big\}.
\eeq
In accordance with ref.\,\cite{bunk_lat94}, we chose $n_{OR}$ to be
roughly equal to the inverse scalar mass in order to achieve maximum
decorrelation. With this choice we found that the average integrated
autocorrelation time estimated using the scalar mass was close to one,
in agreement with \cite{bunk_lat94}.

Our simulations were performed for inverse gauge couplings $\beta_G =
7,\,9 $ and~12. We restricted our attention to one point in the
symmetric and one point in the Higgs region of parameter space
chosen sufficiently away from the phase transition, 
so that the system does
not tunnel between the phases.
In order to compare our results directly with those of a previous
calculation of the lightest scalar and vector masses \cite{ilg95}, 
we work at the same
fixed value of $\lambda_3/g_3^2$,
\beq \label{eq:ratio}
 \frac{\lambda_3}{g_3^2} = \frac{\beta_R\,\beta_G}{\beta_H^2} = 0.0239,
\eeq
which in the context of the four-dimensional theory corresponds to a
Higgs mass at tree-level and zero temperature of $M_H \simeq 35\,\rm GeV$.

In the symmetric phase of the model we initially chose $\beta_G=12$,
$\beta_H=0.3411$. The value of $\beta_R$ was then fixed by the 
relations (\ref{lcp1})-(\ref{lcp3}). The same relations determine the 
continuum scalar mass parameter in the $\overline{MS}$ scheme as
$\mu_3^2/g_3^4=0.089$.
Our point in the Higgs phase of the model was fixed to be
$\beta_G=12,\,\beta_H=0.3418$, which corresponds to 
$\mu_3^2/g_3^4=-0.020$ in the continuum. At $\beta_G=7$ and~9 the corresponding
values of $\beta_H$ and $\beta_R$ were chosen according to the ``lines
of constant physics", eqs.\,(\ref{lcp1})--(\ref{lcp3}), using the
constraint eq.\,(\ref{eq:ratio}). 

At $\beta_G=12$, Monte Carlo runs were performed on lattice sizes
ranging from $10^2\cdot 12$ up to $40^3$ in order to analyse finite-size
effects in detail. This is of special importance in the symmetric
phase of the model, where 
we are particularly interested in the possible occurence of
very light states.

For all our observables, statistics were gathered from about 30\,000
compound sweeps. In a few cases, statistics were increased to a total
of 75\,000 sweeps.

\subsection{Constructing improved operators}
\label{sec:imp}

The main difficulty encountered in recent attempts to compute the
mass spectrum in the symmetric phase of the SU(2) Higgs model
\cite{mon95,ilg95}, was the low signal-to-noise ratio in the
computation of the correlation function 
\bea
  C(t) \equiv  \sum_{{\bf x},{\bf x}'}\,
       {\rm e}^{i{\bf p}\cdot ({\bf x}-{\bf x}')}
       \<\varphi^\dagger({\bf x},t)\varphi({\bf x}',0)\>_c 
 & = & \sum_{n>0}\,|\<0|\varphi(0)|n\>|^2\, {\rm e}^{-aE_n\,t} \\
 & \equiv & \sum_{n>0}\,|c_n|^2\,{\rm e}^{-aE_n\,t}
       \stackrel{t\rightarrow \infty}{\simeq} |c_1|^2\,{\rm e}^{-aE_1\,t},
\label{corr}
\eea
where $\varphi({\bf x},t)$ denotes any one of the operators in
eqs.\,(\ref{ops}), and $E_n>E_{n-1}$ is implied.
For our numerical calculation of the masses 
we use the zero momentum timesclice averages
of the original operators, i.e., ${\bf p}=0$ in the above expression. 

One important goal of our simulations is to investigate the possible
existence of very low-lying states in the symmetric phase of the
model, such as predicted by the analytic approach in \cite{bp94}.
If an operator has a bad projection onto the lightest state
one must be able to follow the signal to sufficiently large $t$
before the ground state dominates $C(t)$. 
A poor signal-to-noise ratio of the correlation function
will then hamper any effort to establish the existence of such a state.
The problem is further
exacerbated in the symmetric phase, where, due to the confining
behaviour of the theory, the particle spectrum may consist
of bound states, having a larger spatial extension than their
point-like counterparts in the Higgs region. Previous experience with
calculations of the glueball spectrum in pure gauge theory shows that
conventional local operators have indeed a bad projection onto bound
states in confining theories \cite{tep87,tep92}. The situation could
be considerably improved by constructing ``blocked" or ``smeared",
non-local operators \cite{tep87,albanese}, which are of similar
extended structure as the bound states they are supposed to project
on. Similar techniques, which preserve gauge invariance, have been
developed and successfully applied in simulations of lattice QCD
\cite{QCD_smearing}. 

Here we are applying and extending these ideas in order to construct
non-local versions of the operators defined in eq.\,(\ref{ops}). Some
of these techniques were applied to the four-dimensional SU(2) Higgs
model in ref.\,\cite{how89}.

Following \cite{tep87}, we construct composite (``blocked") link
variables $U^{(n)}_\mu(x)$ of blocking level $n$ according to 
\bea \label{lbl}
U_{\mu}^{(n)}(x)&=&U_\mu^{(n-1)}(x)U_{\mu}^{(n-1)}(x+\hat{\mu})\\ \nn
&&+\sum_{\nu=\pm 1,\nu\neq\mu}^{\pm 2} 
U^{(n-1)}_\nu(x)U_\mu^{(n-1)}(x+\hat{\nu})
U_\mu^{(n-1)}(x+\hat{\mu}+\hat{\nu})U_\nu^{(n-1)\dagger}(x+2\hat{\mu})\;.
\eea
The links $U^{(n)}_\mu(x)$ are twice as long as those at the lower
blocking level $n-1$. We shall refer to this as ``link blocking" in
the following. It seems natural to design a similar procedure for the
scalar fields $\phi(x)$. A ``site-blocked" scalar field
$\phi^{(n)}(x)$ at blocking level $n$ can be constructed iteratively
from a field at a given lattice site and its covariant connection with
the four nearest neighbours,
\beq \label{sbl}
\phi^{(n)}(x)=\phi^{(n-1)}(x)+\sum_{\mu=1}^{2}\left[
U_\mu^{(n-1)}(x)\phi^{(n-1)}(x+\hat{\mu})+U_\mu^{(n-1)\dagger}(x-\hat{\mu})
\phi^{(n-1)}(x-\hat{\mu})\right]\;.
\eeq
Clearly both blocking procedures can
be iterated, thereby quickly increasing the number of links and sites
contributing to a given composite variable.

Non-local blocked operators are now constructed from the basic ones
(\ref{ops}) by replacing the scalar and link variables with composite
ones at a desired blocking level.  Note that the blocking steps are
constructed in a way which preserves the gauge invariance of the
original operators.  The basic operators $R$ and $P$ contain only site
and link variables, respectively. By applying the corresponding
blocking procedure to these operators, we get $N$ operators of
different spatial extension, where $N$ denotes the maximal blocking
level.  These we write as $R^{(n)}(x)$ and $P^{(n)}$,
with $n=0,...N$.  For the operators $L$ and $V$ both site and link
blocking can be applied, so from each of them we construct a set of
$N\times N$ operators, denoted by $L^{(nm)}_{\mu}(x)$ and
$V^{(nm)}_{\mu}(x)$, $n,m=0,...N$, where the first upper index stands
for site and the second for link blocking.

\subsection{Cross correlations}
\label{sec:cross}

The blocking procedure described in the previous subsection is
designed to yield an optimal operator for a given set of quantum
numbers. In an attempt to further separate the excitations from the
ground state we can also utilise the information contained in our
non-optimal operators by considering cross correlations between
different operators in the same channel.

For a given set of quantum numbers we construct a set of,
say, $N$ lattice operators, $\phi_i: i=1,..,N$,
with those quantum numbers. We normalise these operators
so that $\langle {\phi_i}^\dagger \phi_i \rangle = 1$, and we
impose the same normalisation on all the operators we
discuss below. To find the energy of the
lightest state we use a variational criterion. That is to 
say, we find the linear combination of the $\phi_i$ that
maximises
\beq  \label{c1}
\langle \phi^\dagger(a) \phi(0) \rangle = 
\langle \phi^\dagger {\rm e}^{-Ha} \phi \rangle \; .
\eeq
Call this operator $\Phi_1$. In the limit where the
basis $\{ \phi_i\}$ is complete, this procedure
becomes exact. That is to say, if the lightest state is
$\vert 1 \rangle$ and the corresponding energy is $E_1$,
then
\beq \Phi_1 \vert vac \rangle = \vert 1 \rangle\;,
\eeq
and
\beq \label{c2}
{\rm e}^{-aE_1} = \langle {\Phi_1}^\dagger(a) \Phi_1(0) \rangle\;.
\eeq
We can find higher excited states by a simple extension of
this procedure. Let the first excitation be $\vert 2 \rangle$ 
and let the corresponding energy be $E_2$.
We consider the subspace $\{\phi_i\}^{'}$ of 
$\{\phi_i\}$ that is orthogonal to $\Phi_1$, i.e., such that 
$ \langle {\Phi_1}^\dagger(0) \phi(0) \rangle =0$. We apply the same
variational criterion as above, but restricted to this
subspace. This gives us an operator $\Phi_2$. In the limit
where our original basis becomes complete, we have 
\beq \Phi_2 \vert vac \rangle = \vert 2 \rangle\;,
\eeq
and
\beq {\rm e}^{-aE_2} = \langle {\Phi_2}^\dagger(a) \Phi_2(0) \rangle\;. 
\eeq
We can continue this procedure obtaining operators
$\Phi_3, \Phi_4, ...$ from which we can obtain
the energies of higher excited states.

In our case our basis is finite, and we can obtain at most $N$
operators $\Phi$. With such a limited basis eq.\,(\ref{c2}) provides
at best an estimate for $aE_1$. We improve upon this estimate by
calculating the correlation function $\langle {\Phi_1}^\dagger(t)
\Phi_1(0) \rangle $ for all $t$. If we define an effective energy by
\beq
 {\rm e}^{-aE_{{\rm eff}}(t)t} =  
\langle {\Phi_1}^\dagger(t) \Phi_1(0) \rangle\;,
\eeq
then we know that $E_{{\rm eff}}(t)$ will approach $E_1$ from above as
$t$ increases. The more effective our variational procedure, the
smaller the value of $t$ at which this occurs (for a basis that is
complete we would find $E_{{\rm eff}}(t) = E_1$ for all $t$). So we
can estimate $aE_1$ from the value of $aE_{{\rm eff}}(t)$ on its
`plateau'. In practice, what we actually do is to fit the correlation
function to an exponential in $t$ for $t$ large enough (as described
below). From the exponent we then obtain our estimate for $aE_1$. From
the coefficient of the exponential we obtain the normalised projection
of our operator onto the lightest state, i.e.  $\vert \langle 1 \vert
\Phi_1 \vert vac \rangle \vert ^2$.  If we have a good basis of
operators then this projection will be close to one. In practice this
is always the case in the scalar channel, where the projection is
often consistent with unity. In the vector channel the projection tends
to be $\sim 0.8$.

We follow the same procedure for excited states, extracting $aE_i$ by
fitting an exponential to $\langle \Phi_i(t) \Phi_i(0) \rangle$ for
large enough $t$. One must be more careful here than with the ground
state because, with a finite basis, the operator $\Phi_i$ will have
some projection onto all states, not just onto $\vert i \rangle$. So
as $t \to \infty$ its correlation function will ultimately vary as
$\sim \exp(-E_1 t)$ and not as $\sim \exp(-E_i t)$. Thus, by fitting
an exponential at larger $t$ we may underestimate the value of $aE_i$.
In practice this is not a problem where the operators are very good.
For example if the projection of $\Phi_1$ onto the lightest state is
$1-\epsilon$, then the projection of $\Phi_2$ onto this lightest state
is $\leq \epsilon$. If $\epsilon$ is as small as it is in our
calculations, then this potential contamination of $E_2$ by $E_1$ is
insignificant. The same argument can be used for higher excited
states.  In general, where we quote a mass without qualifications, 
we are confident, by
examining the relevant projections, that our mass estimate is not
significantly contaminated by admixtures of any of 
the lighter states that we list.

In practice our lattice is finite and so in the above we
replace ${\rm e}^{-Et}$ by ${\rm e}^{-Et} + {\rm e}^{-E(T-t)}$ where $T$ is
the length of the lattice in the $t$-direction.

The procedure we follow to obtain the $\Phi_i$ is standard
\cite{matrix_corr}.  Define the $N \times N$ correlation matrix $C(t)$
by
\beq \label{cij}
C_{ij}(t) =  \langle {\phi_i}^\dagger(t) \phi_j(0) \rangle\; .
\eeq
Let the eigenvectors of the matrix $C^{-1}(0) C(a)$ be
$v^i ; i=1,\ldots ,N$. Then
\beq \label{eigen}
\Phi_i = c_i \sum_{k=1}^N v_k^i \phi_k \equiv \sum_{k=1}^N a_{ik}\phi_k
\; ,
\eeq
where the constant $c_i$ is chosen so that $\Phi_i$ is
normalised to unity. 

We would like to emphasise that there are many possible variations on
the above variational procedure. For example we could apply it to
$t=2a$ rather than to $t=a$. As a check we have performed such an
alternative calculation. We further remark that, in practice, the best
of our original $\phi_i$ operators is already so good that the
calculation of the ground state in each channel is not greatly
improved by going to the $\Phi_i$ operators. It is if we wish to
obtain the excited states that this analysis becomes indispensable.

In our actual calculations, we typically compute a $9\times9$
matrix of correlators
in the $0^{++}$ channel, which consists of the three operators $R,\,P$
and~$L$, each taken at three different blocking levels. In the
$1^{--}$ channel, where only operators of type~$V$ are known, three
different blocking levels are used to compute a $3\times3$ correlation
matrix. 

\subsection{Fits and error analysis}

All our mass estimates are obtained from measured correlation
functions of operators $\Phi_i$ in the diagonalised basis defined in
the preceding subsection. The ansatz we use for the asymptotic
behaviour of the correlation function on a finite lattice, for large $T$, is
\beq \label{eq:asymp}
\widetilde{C}_i(t) \equiv \langle\Phi^\dagger_i(t)\Phi_i(0)\rangle
 = A_i\left({\rm e}^{-aM_i\,t} + {\rm e}^{-aM_i(T-t)}\right)\;,
\eeq
where $i$ labels the operator, and $T$ denotes the extent of the
lattice in the time direction.
This expression would be exact for all $t$ if the basis of operators was
complete. To monitor deviations from this behaviour we define an
effective mass according to 
\beq \label{meff}
aM_{\rm eff}(t) = {\rm arcosh}\left\{\frac{C(t+1)+C(t-1)}{2\,C(t)}
                              \right\}\;,
\eeq
where $C(t)$ denotes either $\widetilde{C}_i(t)$ or $C_{ii}(t)$.
As one readily sees, this definition has the desired property that
$aM_{\rm eff}(t)=aM_i$ for those $t$ where $C(t)$ is accurately given by
(\ref{eq:asymp}).

Estimates for the masses $aM_i$ and amplitudes $A_i$ are obtained from
correlated fits of $\widetilde{C}_i(t)$ to eq.\,(\ref{eq:asymp}) over a
finite interval $[t_1,\,t_2]$. Our choice of the fitting interval 
is guided by
the plateaux observed in the effective masses (\ref{meff}),
and is constrained by the requirement that a 
reasonable $\chi^2/\rm dof$ should be obtained.

Our individual measurements of $\widetilde{C}_i(t)$ are accumulated in
bins of typically 500 measurements each. Statistical errors on $aM_i$
and $A_i$ are obtained from a jackknife analysis of the fits to the
average of $\widetilde{C}_i(t)$ in each jackknifed bin.

It has been known for some time that correlated fits may
amplify hidden systematic errors in the data \cite{seibert}. Therefore
we repeated all our fits using an uncorrelated covariance matrix. The
difference between the results obtained using either correlated or
uncorrelated fits are quoted as a (symmetric) systematic error on our
mass estimates. In most cases we found the systematic error arising
from this procedure much smaller than the statistical error. For the
final extrapolation of masses and mass ratios to the continuum limit,
statistical and systematic errors are added in quadrature before the
extrapolation is performed.

\section{Results}
\label{sec:results}

In this section, we present our main results. 
We start with a
discussion of the effects of the blocking procedure and the
diagonalisation of operators in subsections\,\ref{sec:blocking}
and\,\ref{sec:diagonal}, using our data at $\be_G=12$ on the largest
lattices we investigated in the confinement phase ($40^3$), and in the
Higgs phase ($20^3$). The main results on the spectrum, which were
obtained using diagonalised operators at all three values of $\be_G$,
are presented in subsections\,\ref{sec:spectrum}
and\,\ref{sec:excited}. Finally, in subsection\,\ref{sec:cont} we
give our mass estimates extrapolated to the continuum limit.

\subsection{The effects of the blocking procedure}
\label{sec:blocking}

\begin{figure}
\begin{center}
\leavevmode
\epsfysize=250pt
\epsfbox[20 30 620 730]{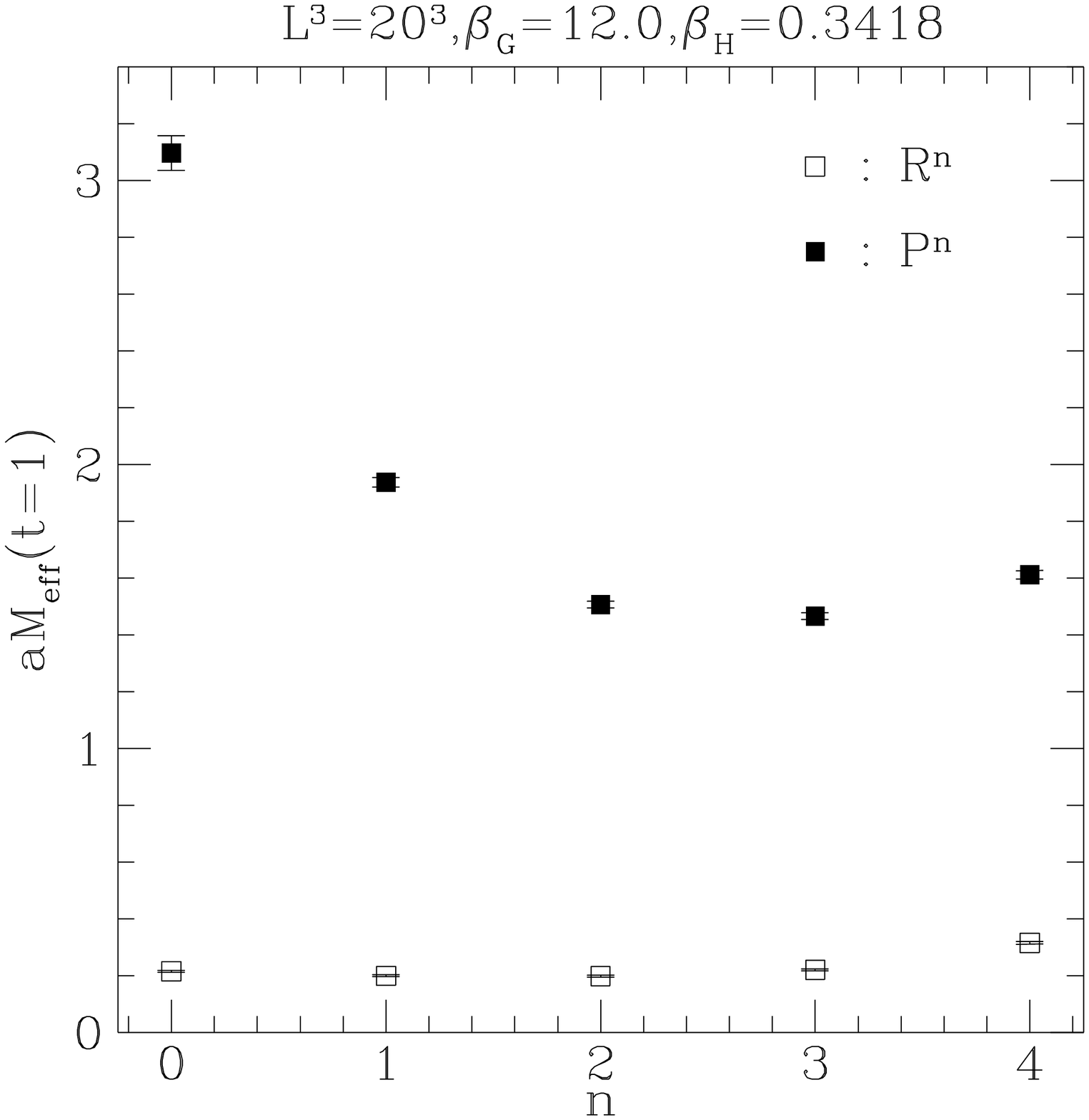}
\leavevmode
\epsfysize=250pt
\epsfbox[20 30 620 730]{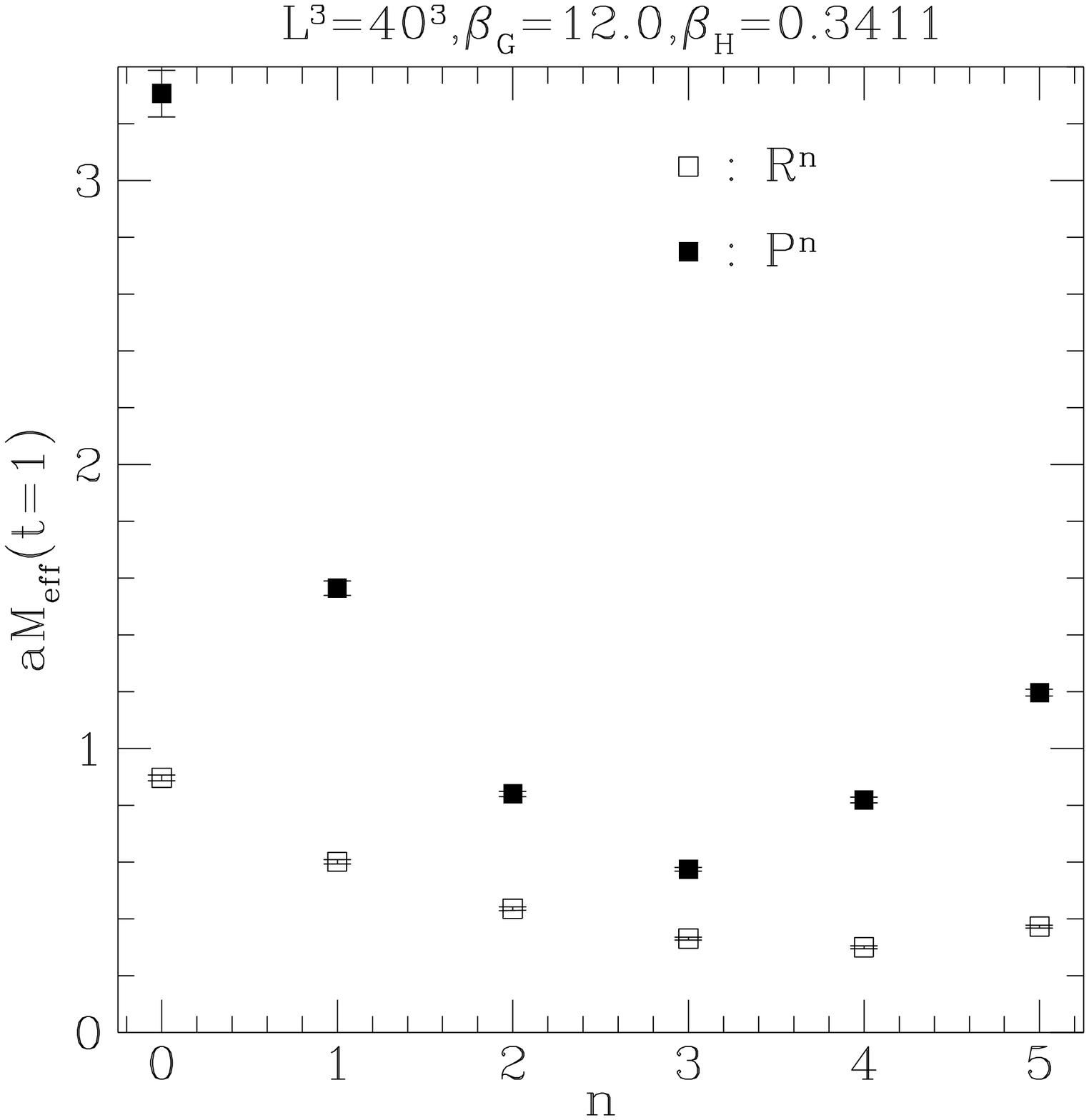}
\vspace{-1.6cm}
\end{center}
\caption[]{\it \label{blo1_2}
 Effect of blocking on the $R$ and $P$ operators in the Higgs (left)
 and confinement (right) phases.}
\end{figure}

A priori nothing is known about the projection properties of the
individual operators in our set $\{\phi_i\}$. The candidates with the
best projection onto the lowest states have to be determined from
actual simulations.  A criterion to judge the performance of an
operator is its effective mass at time separation one, where 
the lowest value indicates the
least contamination from excited states.  Figure~\ref{blo1_2}
illustrates the effect of the blocking procedure for the purely
scalar/gauge operators $R/P$ (cf.~eqs.(\ref{ops})) 
in the Higgs phase and the confinement
phase, respectively.
In the Higgs phase, nothing is
gained by blocking the $R$ operator, while in the
confinement phase four iterations are necessary before it
reaches its optimal projection. For the plaquette $P$, three blocking
steps are required to get to the minimal effective mass in either
phase, but the improvement is far more pronounced in the confinement
phase.

\begin{figure}
\begin{center}
\leavevmode
\epsfysize=250pt
\epsfbox[20 30 620 730]{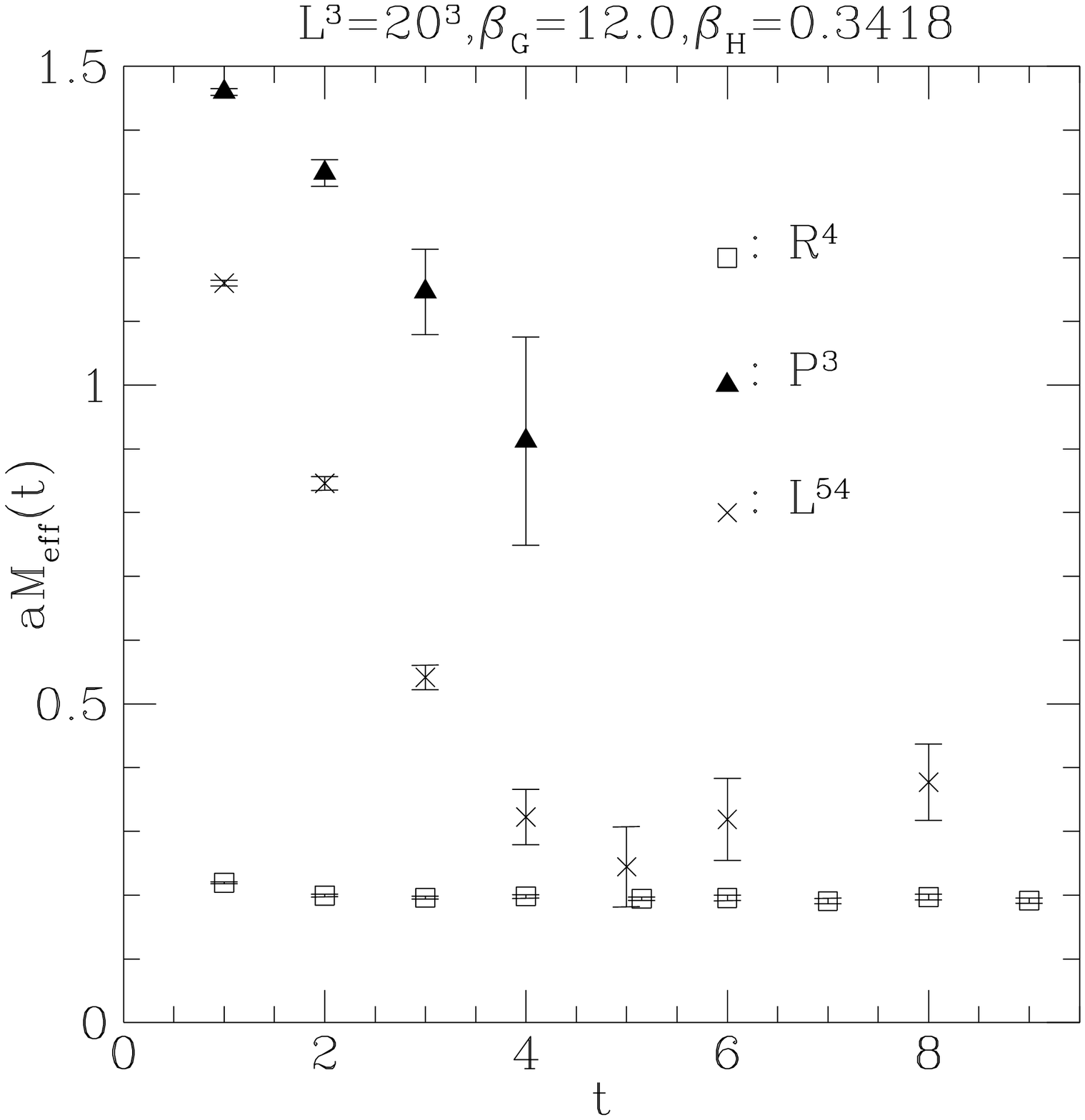}
\leavevmode
\epsfysize=250pt
\epsfbox[20 30 620 730]{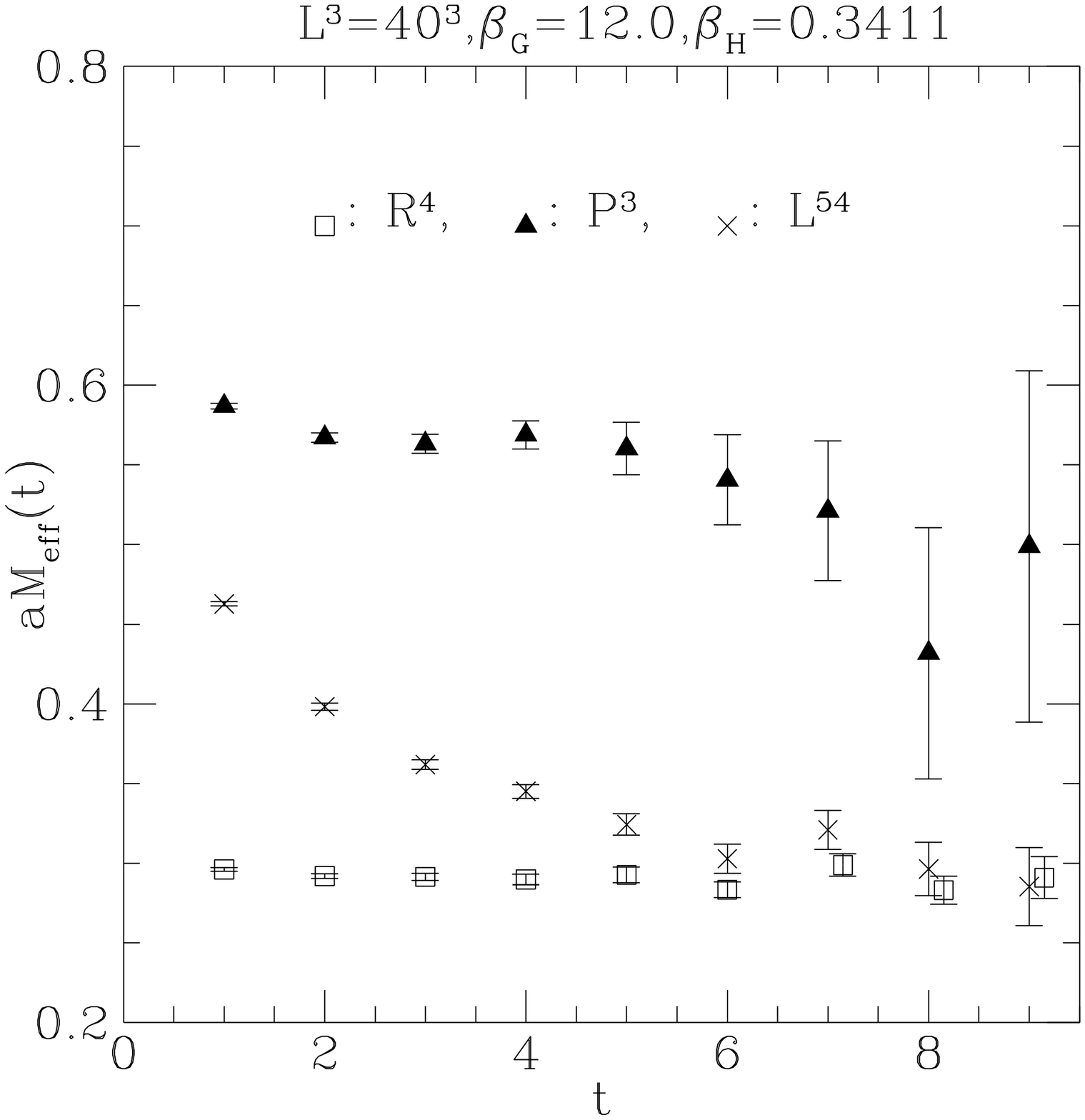}
\vspace{-1.6cm}
\end{center}
\caption[]{\it \label{blo3_4}
 The optimally blocked operator of every basic type in the
 $0^{++}$ channel, Higgs (left) and confinement (right) phase.}
\end{figure}

Figure~\ref{blo3_4} shows effective mass plots for the operators with
the best projection of each basic type in the $0^{++}$ channel, again
for the Higgs and confinement phases, respectively. In both regimes
the operator with the best projection onto the ground state is of the
type $R$, with nearly 100\% overlap at the optimal blocking level. The
ground state could in principle also be extracted from the correlation
function of the best candidate of the type $L$ at large time
separations. However, its projection is much worse, five to six
lattice spacings are needed until excitations have died away, and a
mass calculated from this correlation function would be much less
accurate. Of particular interest is the behaviour of the plaquette
correlations. While they are dominated by noise in the Higgs phase, they
suggest a separate plateau in the effective mass plot in the
confinement phase.  Up to those time separations for which we have a
good signal, there seems to be no tendency for this operator to mix
with the other $0^{++}$ operators. We shall return to this observation
below.

\begin{figure}
\begin{center}
\leavevmode
\epsfysize=250pt
\epsfbox[20 30 620 730]{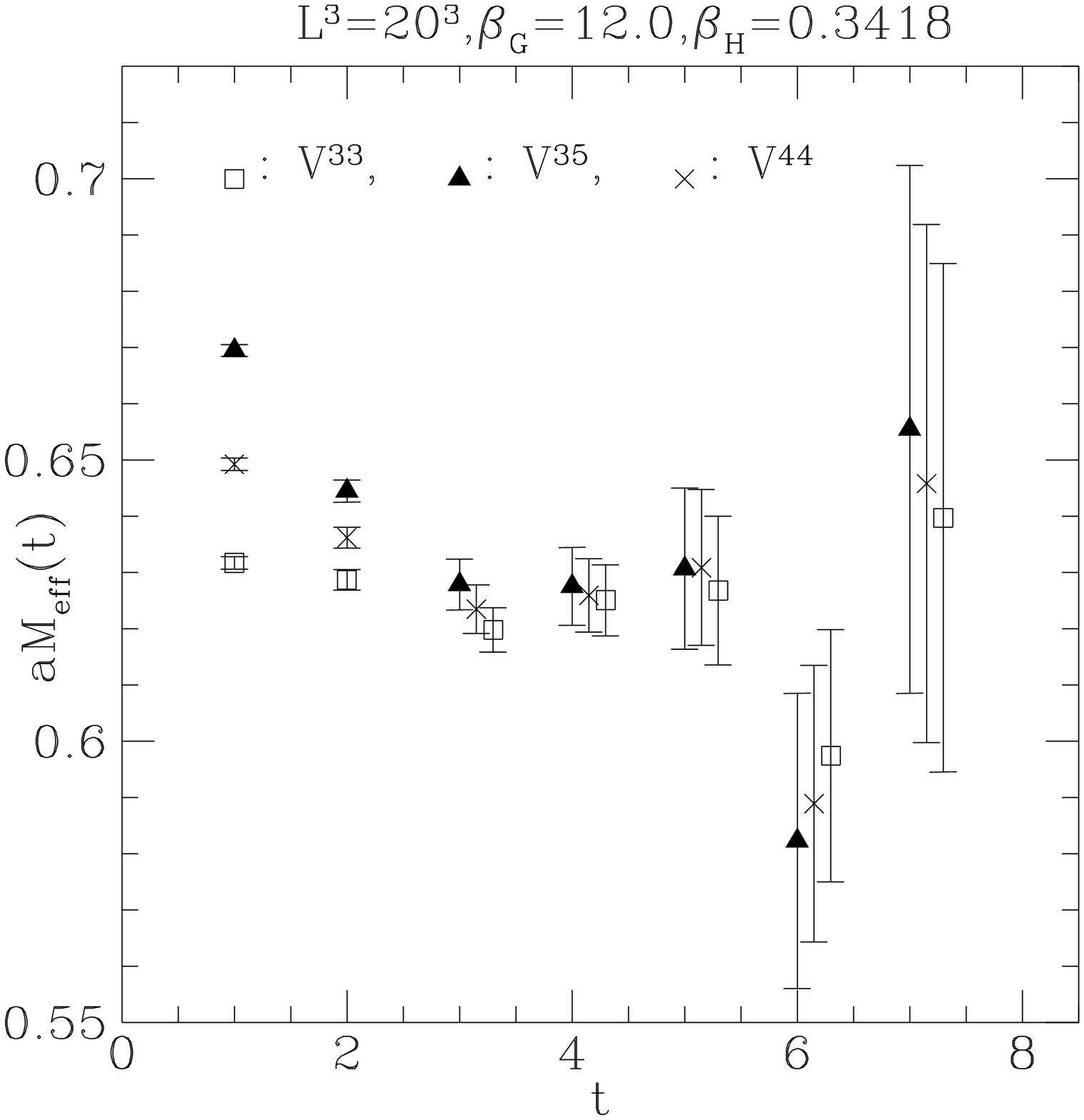}
\leavevmode
\epsfysize=250pt
\epsfbox[20 30 620 730]{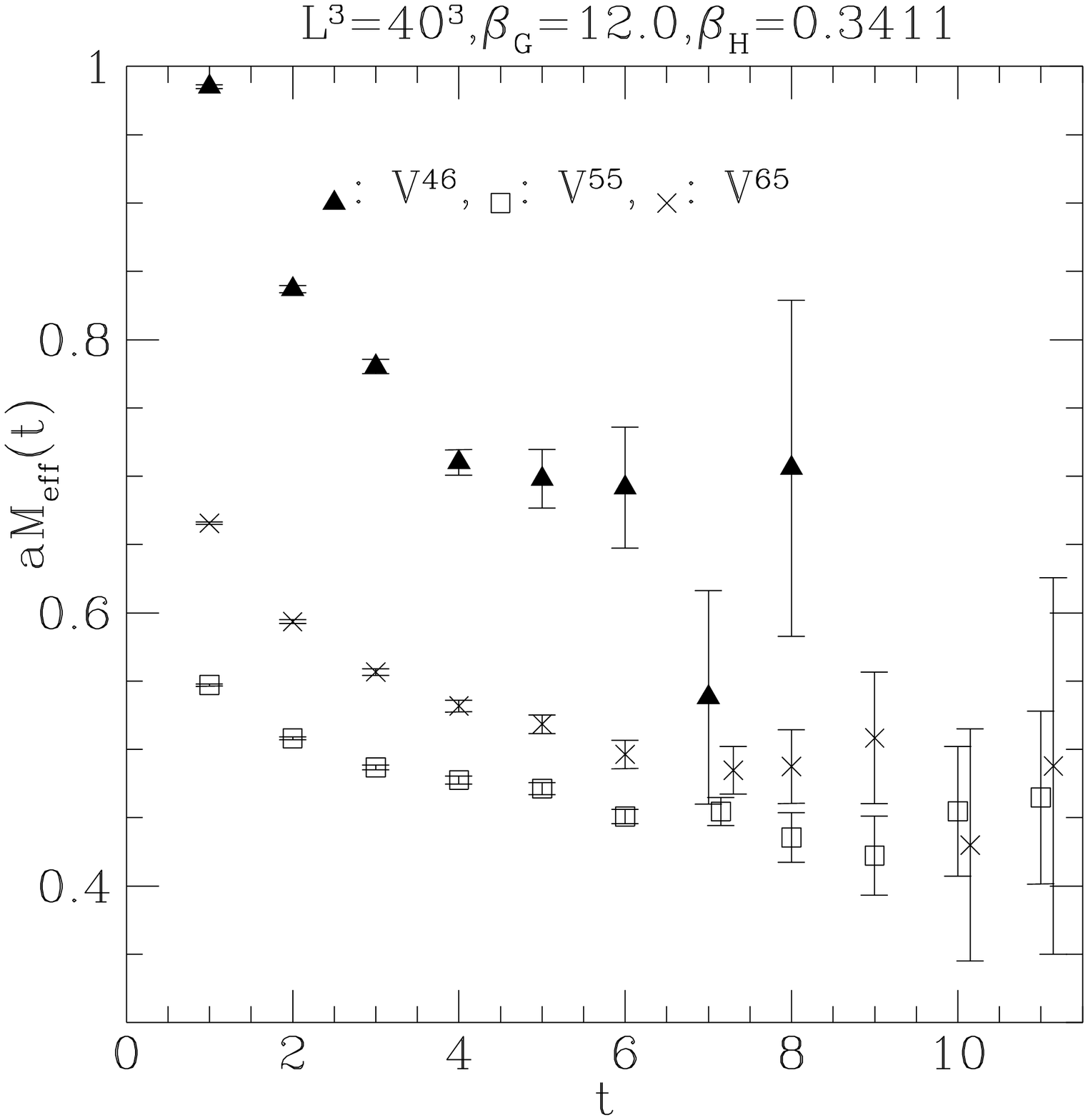}
\vspace{-1.6cm}
\end{center}
\caption[]{\it \label{blo5_6}
 Three differently blocked $W$ operators in the $1^{--}$ channel,
 Higgs (left) and confinement (right) phase.}
\end{figure}

Finally, Figure~\ref{blo5_6} shows the result of the blocking procedure
on the effective masses of the vector boson. In the Higgs phase blocking
slightly improves the projection of the operator $V$, but it is not
difficult to extract a mass also from the unblocked one.  In the
confinement phase the situation is rather different.  The unblocked
operator does not give any signal beyond noise, and the figure
displays how even for the best blocked candidates excitations die out
only very slowly.

In particular, the figure illustrates how one might easily 
extract too large a mass for the vector boson, if one only used a
non-optimal operator such as the one symbolised by the triangles. The
effective masses produced by the different operators do not seem to
merge at a common ground state up to the distances to which we can
follow the signal. An indication that we do really see the
ground state is the fact that further iterations in the
blocking procedures for either links or scalars, beyond the level of
the operator $V^{55}$, again result in a worse projection.
Figure~\ref{blo5_6} also displays a nice side effect of the blocking
procedure. Since the improved operators have a better projection onto
low mass states the corresponding correlation functions fall less
steeply than those of the unblocked ones, hence the signal-to-noise
ratio is improved, leading to considerably smaller statistical errors.

In summary, we find that blocking has little or no effect in the Higgs
phase, where the original local operators exhibit a rather good
projection onto the ground state in each channel.  In the confinement
phase, on the contrary, blocking turns out to be necessary in order to
obtain any useful signal at all. This is particularly pronounced in the
$1^{--}$ channel.  It was also demonstrated by using a large set of
operators, that in the confinement phase 
for time separations up to ten timeslices one is
typically still rather far away from the asymptotic region where all excited
states have died out.  This implies that the latter are rather light
compared to the ones in the Higgs phase, as we shall see more
explicitly in the next subsections. This is precisely what spoils an easy
mass measurement in the confinement phase.  We have increased blocking
levels on each kind of operator until we could explicitly identify the
ones with the best projection onto the lowest states. Thus we can be
sure that we have found the optimal operators that can be constructed from
(\ref{ops}) by means of the blocking technique described in 
subsection\,\ref{sec:imp}.

\subsection{Correlations of eigenstates}
\label{sec:diagonal}

\begin{figure}
\begin{center}
\leavevmode
\epsfysize=250pt
\epsfbox[20 30 620 730]{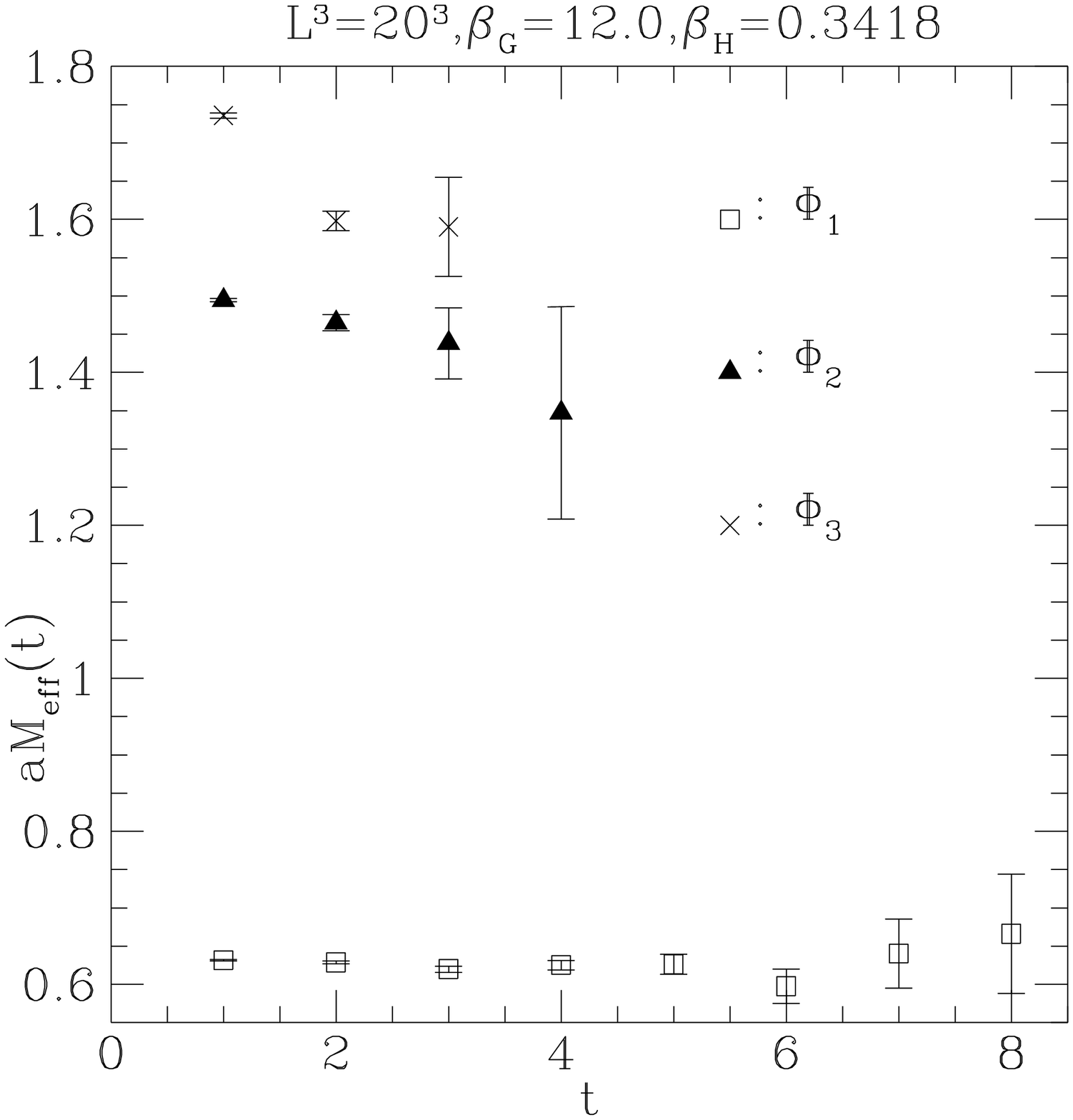}
\leavevmode
\epsfysize=250pt
\epsfbox[20 30 620 730]{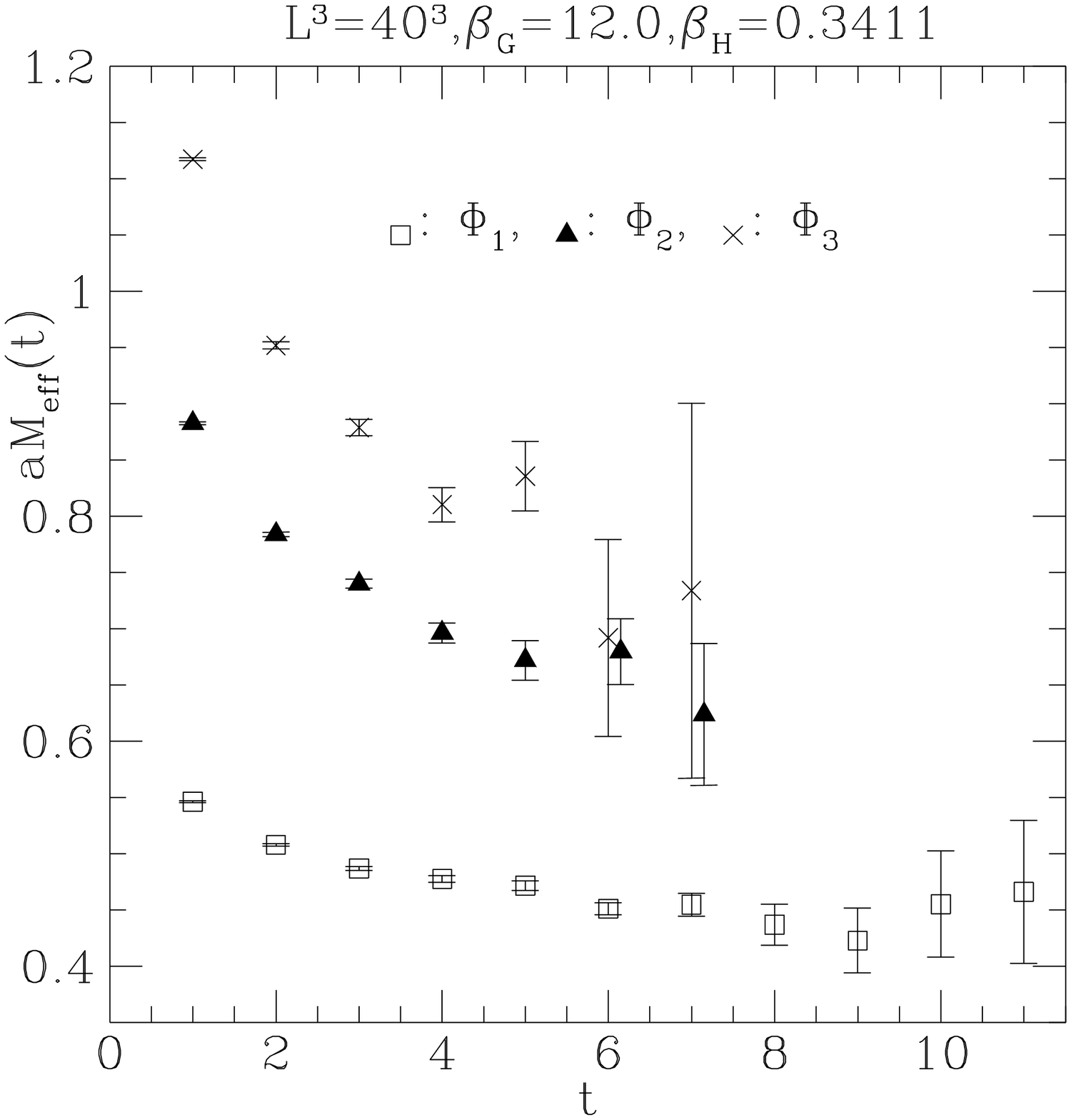}
\vspace{-1.6cm}
\end{center}
\caption[]{\it \label{cross1_2}
 The three lowest eigenstates in the $1^{--}$ channel, Higgs (left)
 and confinement (right) phases.} 
\end{figure}

Now we discuss the correlations of the eigenstates of the matrix
correlators $C_{ij}(t)$ introduced in subsection\,\ref{sec:cross}.
Consider first
the $1^{--}$ channel. Employing a basis composed of the three operators
used in Figure~\ref{blo5_6} we obtain, after diagonalisation,  
the three sets of effective $W$ masses 
shown in Figure~\ref{cross1_2} for the Higgs and
confinement phases.  Comparing with 
Figure~\ref{blo5_6}, we conclude that this
procedure has only slightly improved the projection onto the ground
state. However, it has clearly separated off the excitations. Even
though it is not always possible to identify extended plateaux for these excited
states, one can nevertheless conclude from the comparison between
Higgs and confinement phases that the gap between the ground state and
the excitations is much larger in the former. Clearly, in order to obtain
more quantitative information about the excitations, one would need to
choose a larger basis of operators. We shall not pursue this possibility
here, since in this channel our main interest is in the ground state.

\begin{figure}
\begin{center}
\leavevmode
\epsfysize=250pt
\epsfbox[20 30 620 730]{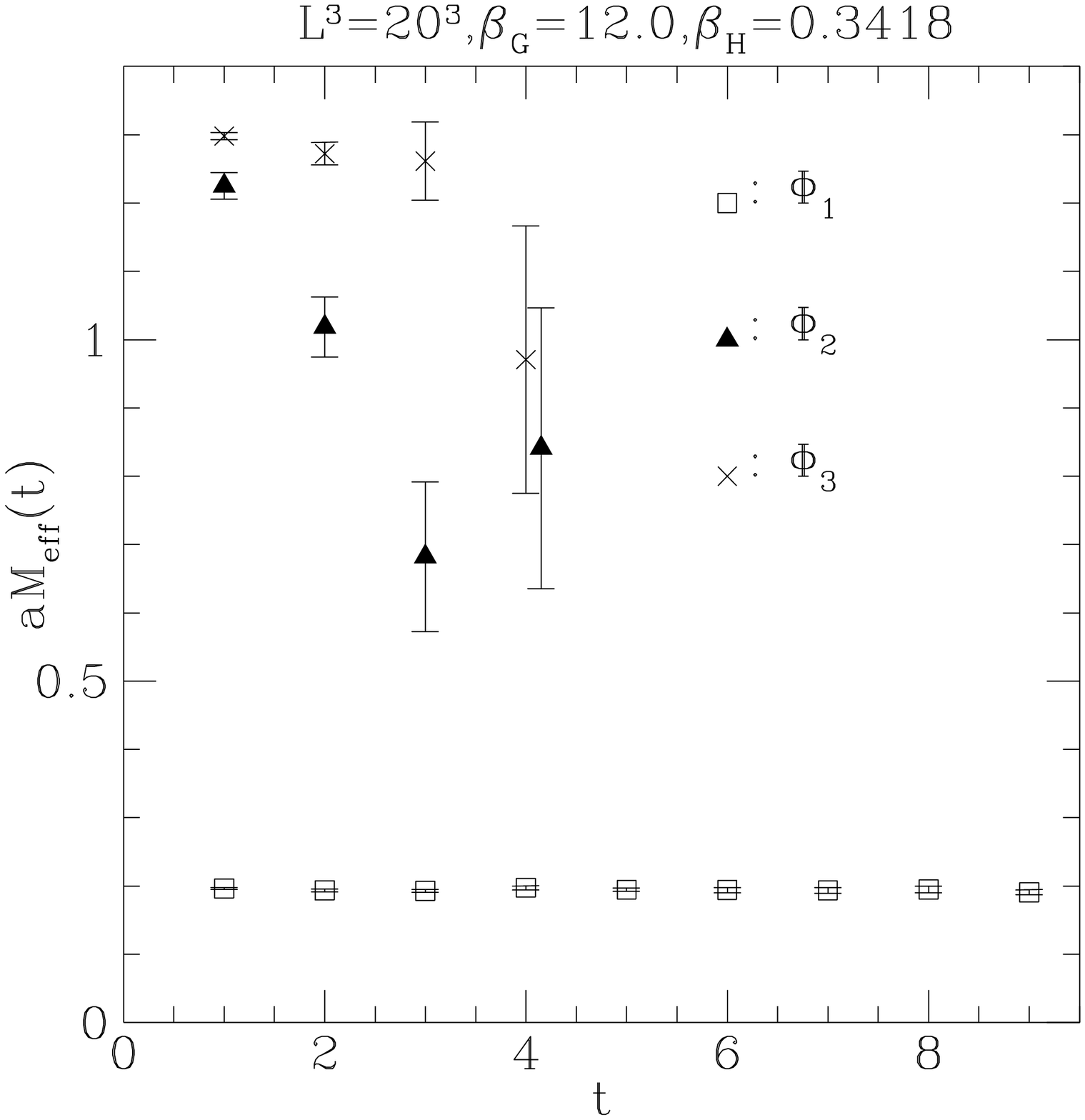}
\leavevmode
\epsfysize=250pt
\epsfbox[20 30 620 730]{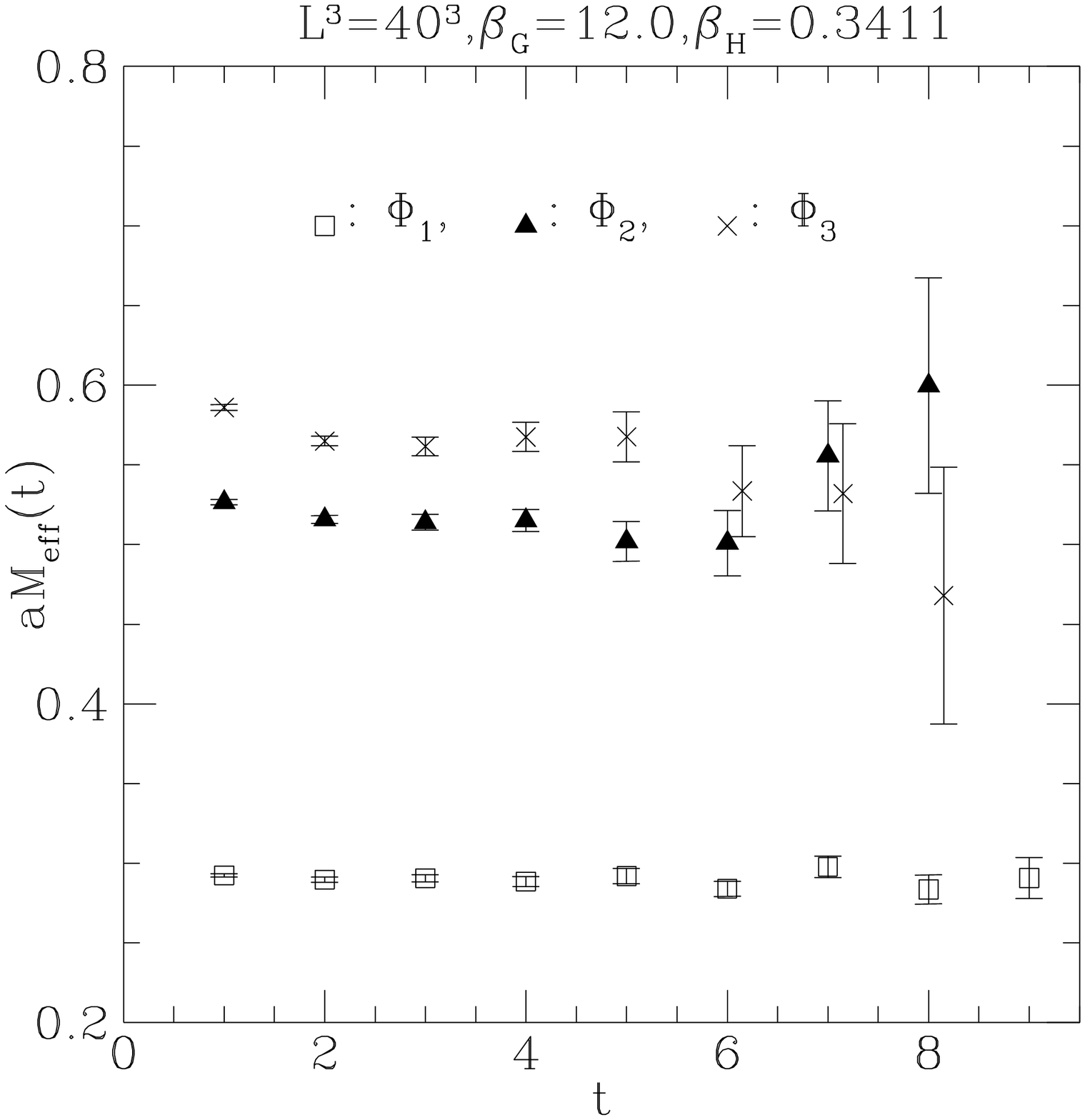}
\vspace{-1.6cm}
\end{center}
\caption[]{\it \label{cross3_4}
 The three lowest eigenstates in the $0^{++}$ channel, Higgs (left)
 and confinement (right) phases.} 
\end{figure}

Next, consider the three lowest states in the $0^{++}$ channel shown 
in Figure~\ref{cross3_4}.  In
the Higgs region, the situation is rather simple
with an isolated Higgs ground state and a large gap to excitations.
Because the excited states are much higher in mass, their correlation
functions fall rapidly, one loses the signal after a few
timeslices, and it is difficult to identify well-defined excited
states. Here one would also need to increase the basis of operators 
and to reduce the lattice spacing, in
order to improve the situation.  In contrast, in the confinement phase
the diagonalisation has isolated three distinct states which were
mixed previously. In Table \ref{coeff} the coefficients $a_{ij}$ 
(cf.~eq.~(\ref{eigen})) with
which the individual operators contribute to various eigenstates are
shown.  The labelling is such that $\Phi_1$ denotes our best operator 
for the ground state, $\Phi_2$ the one for the first excited state, etc.

\renewcommand{\arraystretch}{1.5}
\begin{table}[ht]
\begin{center}
\begin{tabular}{||c|r@{.}lr@{.}lr@{.}lr@{.}lr@{.}l||}
\hline
\hline
  & \multicolumn{2}{c}{$\Phi_1=H1$}
  & \multicolumn{2}{c}{$\Phi_2=H2$} 
  & \multicolumn{2}{c}{$\Phi_3=H3$}
  & \multicolumn{2}{c}{$\Phi_6=H3^*$}
  & \multicolumn{2}{c||}{$\Phi_9=H3^{**}$} \\
\hline
$\phi_1=R^3$   & 0&932(1) & 0&315(4) & 0&034(7) & 0&0007(54) & 0&0019(13)\\
$\phi_2=R^4$   & 0&9962(2) & 0&017(4) & 0&024(3) & 0&0020(53) & 0&0018(11)\\
$\phi_3=R^5$   & 0&850(2) & 0&493(4) & 0&07(1)  & 0&0029(50) & 0&0041(16) \\
$\phi_4=P^2$   & 0&071(3) & 0&11(1)  & 0&615(3) & 0&64(1)    & 0&435(8)\\
$\phi_5=P^3$   & 0&068(3) & 0&19(2)  & 0&973(4) & 0&017(4)   & 0&107(2)\\
$\phi_6=P^4$   & 0&036(3) & 0&13(1)  & 0&631(4) & 0&594(9)   & 0&476(8)\\
$\phi_7=L^{44}$& 0&818(2) & 0&200(5) & 0&048(6) & 0&01(4)    & 0&005(7)\\
$\phi_8=L^{54}$& 0&661(3) & 0&562(4) & 0&05(1)  & 0&004(28)  & 0&012(5)\\
$\phi_9=L^{65}$& 0&337(3) & 0&655(4) & 0&021(13)  & 0&08(7)    & 0&04(2)\\
\hline
\hline
\end{tabular}
\end{center}
\caption[]{\it Coefficients $a_{ij}$ as defined in
  eq.\,(\protect\ref{eigen}) of the operators used in the simulation
  for the three lowest $0^{++}$ states in the confinement phase  
  ($\beta_G=12$, $\beta_H=0.3411$, $L^2\cdot T=40^3$). In the header, we also
  introduce the labelling for scalar states used below.\label{coeff}
}
\end{table}

According to this analysis the ground state in the confinement phase
consists predominantly of
$R$- and $L$-contributions.  The next state has
contributions from all types of operators, with a dominance by
$R$ and $L$. As in the spin-one case, the gap between the lowest and
first excited states is much smaller than in the Higgs phase. The separate
plateau of the plaquette operators survives diagonalisation,
representing a rather definite state. Table \ref{coeff} shows that the
plaquette operators indeed have practically no overlap with the ground
state $\Phi_1$. Conversely, the other $0^{++}$ operators do not
contribute to the state $\Phi_3$, which thus appears to be an object
composed exclusively of gauge degrees of freedom and very little
mixing with operators containing scalars. In the pure gauge theory
this object would correspond to a glueball. It seems natural to
interpret this state in the Higgs model as a $0^{++}$ ``$W$-ball",
composed almost entirely of gauge bosons.  As shown in Table \ref{coeff}, the
basis of eigenstates contains two more states $\Phi_6,\Phi_9$ with
almost exclusively plaquette contributions, thus appearing to be
excitations of the state $\Phi_3$. Details of the spectrum of excited
states will be presented in subsection\,\ref{sec:excited}.

\subsection{Mass spectrum and finite-size analysis}
\label{sec:spectrum}

Now we proceed to presenting our complete set of results for the
spectrum of the SU(2) Higgs model in three dimensions. We perform an
analysis of finite-size effects and finally extrapolate our results to
the continuum limit.

In Tables \ref{results_hm} and \ref{results_cm} we summarise the
results on all lattices and for all values of $\beta_G,\,\beta_H$
used in our calculation.  All masses quoted in this section have been
obtained by fitting the correlation functions to the functional form
in eq.\,(\ref{eq:asymp}).  As has been demonstrated in 
subsection\,\ref{sec:diagonal}, 
our signals for the lowest states show quite pronounced
plateaux. The situation is more difficult for the excited states.  In
Tables \ref{results_hm} and \ref{results_cm} we only record masses for
which we could identify a plateau of at least three timeslices
extension in an effective mass plot. Those cases where the statistical
errors of the correlation function were large, or where the overlap
$a_{ij}$ of the diagonalised operators onto the desired state was small,
are marked by an asterisk.

\begin{table}[tbhp]
\renewcommand{\baselinestretch}{1.}
\begin{center}
\begin{tabular}{||r|r@{.}l|l||r@{.}l|r@{.}l|r@{.}l||}
\hline
\hline
$\beta_G$
& \multicolumn{2}{c|}{$\beta_H$}
& $L^2\cdot T$
& \multicolumn{2}{c|}{$aM_{H1}$} 
& \multicolumn{2}{c|}{$aM_{H3}$} & \multicolumn{2}{c||}{$aM_W$} \\
\hline
\hline
12 & 0&3418  & $20^3$        & 0&1944(13)(1)  
            & 1&270(16)(5)$^*$   & 0&624(5)(1) \\
   &\multicolumn{2}{c|}{} & $16^2\cdot32$ & 0&1955(8)(9) 
   &\multicolumn{2}{c|}{} &  0&625(2) \\
\hline
 9 & 0&3450 & $14^2\cdot20$ & 0&2627(14)(4)  
            & 1&90(5)$^*$   & 0&836(2)(1) \\
\hline
 7 & 0&3488 & $20^3$        & 0&348(2)(2) 
            &\multicolumn{2}{c|}{}  & 1&067(4)(3) \\
\hline
\hline
\end{tabular}

\caption{\it Mass estimates in the $0^{++}$ and in the $1^{--}$ 
  channels in the Higgs phase. The first error is statistical, the
  second is an estimate of systematic effects.}

\label{results_hm}
\end{center}
\end{table}

\begin{table}[tbhp]
\renewcommand{\baselinestretch}{1.}
\begin{center}
\begin{tabular}{||r|r@{.}l|l||r@{.}l|r@{.}l|r@{.}l|r@{.}l||}
\hline
\hline
$\beta_G$
& \multicolumn{2}{c|}{$\beta_H$}
& $L^2\cdot T$ 
& \multicolumn{2}{c|}{$aM_{H1}$} & \multicolumn{2}{c|}{$aM_{H2}$} 
& \multicolumn{2}{c|}{$aM_{H3}$} & \multicolumn{2}{c||}{$aM_W$} \\
\hline
\hline
12 & 0&3411 & $40^3$        & 0&2903(15)(12) & 0&514(4)(1) 
            & 0&563(5)(2)    & 0&447(8)(3) \\
   &\multicolumn{2}{c|}{} & $32^3$        & 0&2885(25)(15) & 0&440(9)(4)
            & 0&527(13)(13)  & 0&442(4)(5) \\
   &\multicolumn{2}{c|}{} & $26^3$        & 0&2813(22)(5)  & 0&334(10)(4)
            & 0&544(12)(2)   & 0&443(9)(2) \\
   &\multicolumn{2}{c|}{} & $20^3$        & 0&226(6)      & 0&306(8)(3)
            & 0&509(11)(10)   & 0&423(10)(4)\\
   &\multicolumn{2}{c|}{} & $16^2\cdot32$ & 0&1739(32)(19) & 0&247(12)(5)
            & 0&540(7)(3)    & 0&422(8)(4) \\
   &\multicolumn{2}{c|}{} & $10^2\cdot30$ & 0&121(4)(1)    & 0&136(8)(1)$^*$
            & \multicolumn{2}{c|}{}  & 0&469(11)(1)\\
\hline
 9 & 0&3438  & $26^3$        & 0&387(2)(4) & 0&677(14)(3)
            & 0&772(10)(5)    & 0&610(4)(2) \\
\hline
 7 & 0&3467 & $30^3$        & 0&510(2)(2)    & 0&912(12)(4)$^*$
            & 1&003(6)(2)    & 0&799(8)(6)    \\
   &\multicolumn{2}{c|}{} & $20^3$        & 0&512(3)(3)    & 0&908(12)(18)
            & 0&997(20)(3)   & 0&801(4)(2) \\
\hline
\hline
\end{tabular}
\end{center}
\caption{\it Mass estimates in the $0^{++}$ and in the $1^{--}$ 
  channels in the symmetric phase. The first error is statistical, the
  second is an estimate of systematic effects.}
\label{results_cm}
\end{table}

We investigate finite volume effects in detail for $\be_G=12$. 
Numerically, the infinite volume limit is
reached when the change in a mass with increasing lattice size is
smaller than the statistical errors.  
In order to avoid additional finite size studies for the
smaller values of $\beta_G$, we take the required spatial length
corresponding to the large volume limit of the lattice at $\be_G=12$
in units of the Higgs mass, $M_{H1}L$, and scale it down to the lower
$\be_G$-values.  This way we ensure that the simulations at the
smaller values of $\be_G$ are done in the same physical volume as for
$\be_G=12$. Strictly speaking, this procedure is only valid if the
considered range of values for $\be_G$ is in the scaling region, an
assumption which turns out to be satisfied rather well, as we shall
see {\it a posteriori\/}. 
After infinite-volume masses
have been determined for different $\beta_G$-values they can be
extrapolated to $\beta_G\rightarrow\infty$.

The large-volume limit of the Higgs phase is rather easy to reach.
Table \ref{results_hm} gives the Higgs and $W$ boson masses in lattice
units as measured on lattices with spatial lengths $L=16,L=20$ at
$\be_G=12$. It is seen that for both states the masses on the two
lattices are compatible within the statistical errors.

\begin{figure}[tbp]
\begin{center}
\leavevmode
\epsfysize=400pt
\epsfbox[20 30 620 730]{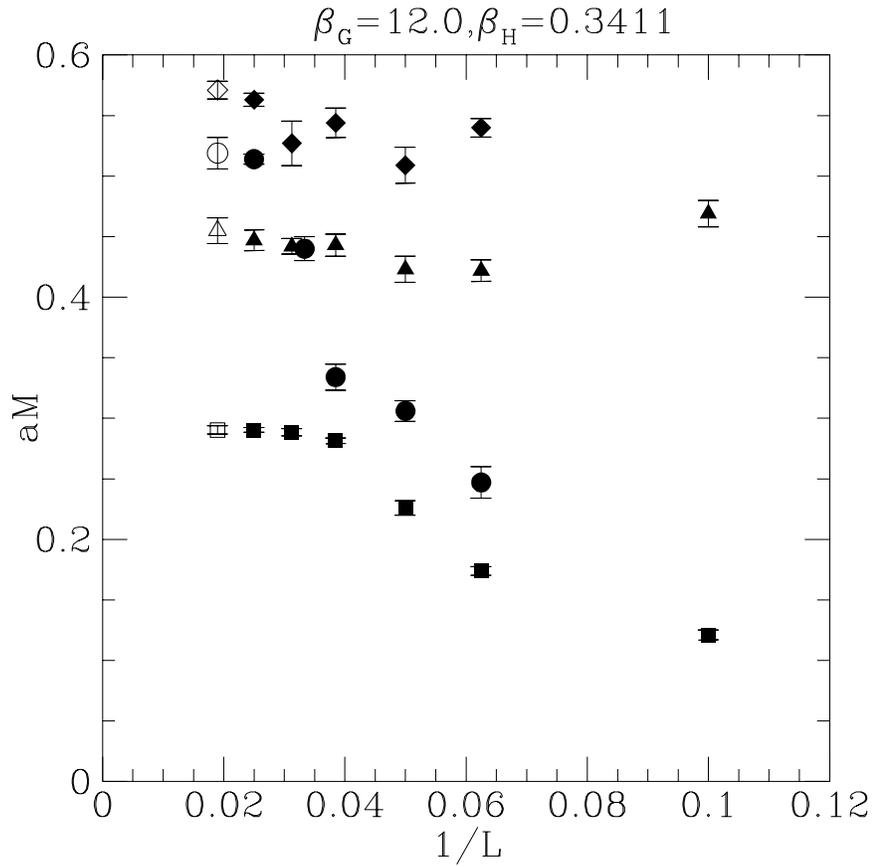}
\vspace{-2.6cm}
\end{center}
\caption{\it Finite volume study for the confinement phase at
  $\be_G=12$. Squares, circles and diamonds represent the three lowest
  $0^{++}$ states, whereas triangles denote the lowest $1^{--}$
  state. Open symbols indicate the data extrapolated from $\be_G=7$
  according to eq.\,(\protect\ref{extra7_12}).
}
\label{vol}
\end{figure}

Again the situation is much more difficult in the confinement phase,
as is illustrated in Figure~\ref{vol}. There are strong finite size
effects for the lightest scalar state, which are only under
control for lattices larger than $L=32\;(1/L=0.031)$. 
We estimate that the
infinite-volume limit for the scalar ground state in the confinement
phase is reached for $aM_{H1}L\simeq10$. 
The vector boson mass, on
the other hand, shows only little dependence on the volume.  The
$W$-ball is just getting close to the large volume limit on a lattice
with $L=40\;(1/L=0.025)$.  The most pronounced finite size effects of
all states investigated are displayed by the intermediate state
$\Phi_2$.  We conclude that the ground state masses in both channels
have reached the infinite volume limit, while for the excitations it
would be desirable to go to larger lattices.  In order to get an
estimate for the situation in larger volumes it is instructive to
consider $\be_G=7$, where a given lattice size 
(here we consider $L=30$) corresponds to a larger
volume in physical units than at $\be_G=12$. Since the mass of the
lowest state $\Phi_1$ is free of finite size effects at $L=40$ 
the ratio of this
mass at the two $\be_G$-values may be used to scale lattice size and
masses according to
\bea
La(\be_G=12)&=&30a(\be_G=7)\frac{aM_{1}(\be_G=7,L=30)}
{aM_{1}(\be_G=12,L=40)}=52.9a(\be_G=12) \;,\nn\\
\quad 
aM_{i}(\be_G=12,L=52.9)&=& aM_{i}(\be_G=7,L=30)
\frac{aM_{1}(\be_G=12,L=40)}{aM_{1}(\be_G=7,L=30)}\;.
\label{extra7_12}
\eea
The result of this scaling is shown as the open data points in
Figure~\ref{vol}. Now the state $\Phi_2$ also seems to have approached
the large volume limit. However, since it is very close to the $W$-ball
at large volumes one expects some mixing between these states.
Comparing our data for the coefficients $a_{ij}$ from the lattices
with $L\ge 26$ we find growing admixtures of plaquette operators to
$\Phi_2$ with increasing volume, while the composition of the state
$\Phi_3$ remains rather unchanged by the approaching $\Phi_2$.  In order
to be absolutely sure that $\Phi_2$ really represents an isolated state further
investigations are required on larger lattices or at different
parameter values, where $\Phi_2$ and $\Phi_3$ might be more clearly
separated.

In ref.\,\cite{ilg95} it was stated that at $\be_G=12$ the results for
the lowest $0^{++}$ state were practically indistinguishable on
lattices of size $30^3$ and $20^3$. In contrast to this, we find a
rather strong dependence of $aM_{H1}$ in this range of lattice sizes. In
particular, our results on $32^3$ and $20^3$ are clearly incompatible.
We ascribe this to a presumably incomplete isolation of the
ground state in ref.\,\cite{ilg95}. 
In addition, we observe that on the $20^3$ lattice
our vector boson mass in the confinement phase is about 35\%, and the 
scalar ground state about 25\% lower than
those reported in \cite{ilg95}. 

We conclude that the construction of improved operators is an
indispensable tool in the study of the mass spectrum of our model in
the confinement phase. In view of this, it would be very interesting
to apply this technique to mass calculations in the vicinity of the
phase transition, and at higher values of $\la_3/g_3^2$ corresponding 
to more realistic zero temperature Higgs masses.

The finite volume effects that we have analysed so far are to do
with the size of the spatial volume. 
There are, however, additional finite volume effects which
have to do with the finite extent in time of the lattice.
Of course the eigenvalues of the transfer matrix and (lattice)
Hamiltonian, $H$, are not altered by varying $T$.
However, what may change is the relationship between those
eigenvalues and the exponents in the decays of our calculated 
correlation functions. For example, the fact that the values 
of our masses are with respect to the vacuum energy arises
because our expectation values contain the partition 
function, $Z = \tr\{e^{-HT}\}$, as a normalisation factor
and this factor will normally be dominated by
the vacuum energy. If $T$ is sufficiently small, however, then
$Z$ may receive significant contributions from excited states
and the masses we calculate from our correlation functions may 
be shifted by the corresponding shift in the `effective'
vacuum energy. Exactly what the effect of this is going to
be is a complicated matter since, on the one hand, similar
contributions occur in the numerator of the correlation function
and this may lead to a partial cancellation of this correction.
On the other hand our scalar masses involve the subtraction
of a vacuum expectation value of the operator, and this will
also be affected. Nonetheless, although we cannot easily estimate
where such effects may be important, we note that since 
the leading correction to $Z$ is $O(e^{-aM_{H1}T})$, we need
to be concerned once $aM_{H1}$ is small.

To obtain a quantitative control over this potential problem,
we have taken our $10^2$ spatial lattice at $\beta_G =12$ (because
it has the smallest value of $aM_{H1}$ and we have repeated the
calculations, with the same basis of operators (which, unusually,
happened to be 6 in this case), for $T=20$ and $T=12$. We
have extracted masses in the same way as on the $T=30$ lattice
and have found $aM_{H1}=0.125(3)$ for $T=20$ and $aM_{H1}=0.111(3)$
for $T=12$. Thus there are no finite-$T$ effects within these small
errors down to $T=20$ and even at $T=12$, where $e^{-aM_{H1}T}
\simeq 0.24$, the shift in the extracted mass is only $\sim 10\%$.
At $T=20$, $aM_{H1}T \sim 2.5$ and this gives us a benchmark 
value for judging when we should be safe from such corrections.
We find no significant effects, within errors, for our other states. 
Of course, these effects may be somewhat different 
in the different phases, and, to the extent that scaling is
violated, at different $\beta_G$. However the volumes that
we use for extracting our final masses have values of $aM_{H1}T$ so
much larger than the above benchmark value that we saw no
reason to repeat this analysis in those other cases.

\subsection{Higher excitations}
\label{sec:excited}

The diagonalisation procedure also enables us
to compute masses of more highly excited states, which were not 
mentioned in Table~\ref{results_cm} . These, however, are
determined with much less accuracy, since the variational basis for
these states is smaller. We nevertheless find it instructive to give a
qualitative discussion of that part of the spectrum. Since the gap
between the ground state and excited states in the Higgs phase is
rather large for both the Higgs and the vector boson (as can be seen, e.g.
in Figures\,\ref{cross1_2} and \ref{cross3_4}), we restrict our
discussion to the symmetric phase.

In Table\,\ref{exc_symm} we present the results for those states where
we felt confident enough to quote a mass estimate. For the $W$-ball, the
correlation functions of the first and second excited states were
those where plaquette contributions were clearly dominant (see e.g. Table
\ref{coeff} for the overlaps of operators $\Phi_6$ and $\Phi_9$ at
$\be_G=12$, $L^2\cdot T=40^3$). 

\begin{table}
\begin{center}
\begin{tabular}{||r|r@{.}l|l||r@{.}l|r@{.}l|r@{.}l||}
\hline
\hline
$\beta_G$
& \multicolumn{2}{c|}{$\beta_H$}
& $L^2\cdot T$
& \multicolumn{2}{c|}{${aM_W^*}$}
& \multicolumn{2}{c|}{${aM_{H3}^*}$}
& \multicolumn{2}{c||}{${aM_{H3}^{**}}$} \\
\hline
\hline
12 & 0&3411 & $40^3$        & 0&682(24) & 0&840(24)
            & 1&02(2)   \\
   &\multicolumn{2}{c|}{} & $32^3$        & 0&622(19) & 0&804(10)
            & 0&983(16)  \\
   &\multicolumn{2}{c|}{} & $26^3$        & 0&636(23) & 0&773(12)
            & 0&974(30)   \\
\hline
 9 & 0&3438  & $26^3$        & 0&854(34)  & \multicolumn{2}{c|}{}
            & \multicolumn{2}{c||}{}    \\
\hline
 7 & 0&3467 & $30^3$        & 1&193(17)  & \multicolumn{2}{c|}{}
            & \multicolumn{2}{c||}{}   \\
   &\multicolumn{2}{c|}{} & $20^3$ & 1&219(21)  & \multicolumn{2}{c|}{}
            & \multicolumn{2}{c||}{}  \\
\hline
\hline
\end{tabular}
\caption{\it Mass estimates for the excitations of the vector boson and the
  $W$-ball in the symmetric phase.}
\label{exc_symm}
\end{center}
\end{table}

It is instructive to compare the mass estimates for the $W$-ball and
its excitations with the glueball spectrum in the pure gauge theory.
In Table \ref{glue} the masses in lattice units of these states are
compared with those of the corresponding $0^{++}$ glueball and its
first two excitations at $\be_G=12$ in three-dimensional pure SU(2)
gauge theory \cite{tep92,tepunp}. The striking agreement between
these states in the two theories indicates a remarkably complete
decoupling of the pure gauge sector from the Higgs part in the SU(2)
Higgs model.

\renewcommand{\arraystretch}{1.5}
\begin{table}[ht]
\begin{center}
\begin{tabular}{||l|r@{.}lr@{.}lr@{.}l||}
\hline
\hline
  & \multicolumn{2}{c}{$aM$}
  & \multicolumn{2}{c}{$aM^*$}
  & \multicolumn{2}{c||}{$aM^{**}$} \\ \hline
SU(2) pure gauge & 0&563(5) & 0&805(8) & 0&982(14)\\ \hline 
SU(2) Higgs      & 0&563(5) & 0&840(24)& 1&02(2) \\
\hline
\hline
\end{tabular}
\end{center}
\caption[]{\label{glue} \it 
Comparison of \,$0^{++}$ glueball and $W$-ball and their first two excitations
at $\be_G=12$.}
\end{table}
 
The existence of a separate $W$-ball which does not mix at all with
other $0^{++}$ states is rather unexpected in view of the coupling
between scalar and link variables in the tree-level action, and
this suppression of mixing must be of dynamical origin. It would be
interesting to see whether this isolation of the pure gauge sector
persists also in the $1^{--}$ channel.

\subsection{How certain is the ground state?}

Measuring and diagonalising the correlation matrices provided us with
valuable insight into the excitation spectrum of the theory.  What can
we say about the existence of very light states?  In the
$0^{++}$ channel we have a nearly complete projection onto the lowest
state, and in the $1^{--}$ case the projection onto the lowest state
looks quite acceptable as well. In the last section it was
demonstrated that our operator basis includes the {\it optimal}
operators which can be obtained from the operator types (\ref{ops}) by
means of the blocking techniques (\ref{lbl}),(\ref{sbl}). Although we
do find the lowest $1^{--}$ state to be about 30--40\% smaller than
that quoted in \cite{ilg95},
our lowest masses are
still much larger than the Higgs and vector boson masses predicted by
the gap equations \cite{bp94}.

Of course, we cannot strictly rule out the existence of a lighter
state which may only show up at distances larger than those up to
which we have a good signal. If there were such states, however, they
would have to have a rather poor overlap with our operator basis.
This can be made more quantitative as follows. All effective
masses presented so far were obtained from (\ref{meff}) under the
assumption that the corresponding correlation function is dominated by
a single lowest state. Let us now assume that there is one lighter
state in each channel such that our measured correlation functions
would correspond to a superposition of two states,
\beq \label{2ex}
\widetilde{C}_i(t)\simeq A_i
    \left( {\rm e}^{-aM_i\,t}+{\rm e}^{-aM_i\,(T-t)}\right)
         + A_0 \left( {\rm e}^{-am\,t}+{\rm e}^{-am\,(T-t)}\right)\;,
\eeq
where $am$ corresponds to the supposed very light mass and $aM_i$ is
of the size of the mass we extracted assuming a single exponential
correlation function as in eq.\,(\ref{eq:asymp}). Fixing the assumed
light mass $am$ to values motivated by the study in \cite{bp94}, we
try to fit our data for the low states by the correlation function
(\ref{2ex}). Some results are presented in Table \ref{2exfit}.

\renewcommand{\arraystretch}{1.5}
\begin{table}[ht]
\begin{center}
\begin{tabular}{||c|r@{.}lr@{.}lr@{.}lr@{.}l||}
\hline
\hline
$am=0.1$
& \multicolumn{2}{c}{$aM_i$}
& \multicolumn{2}{c}{$A_i$}
& \multicolumn{2}{c}{$A_0$}
& \multicolumn{2}{c||}{$\sqrt{(A_0+2\sigma)/A_i}\leq $}  \\ \hline
$\Phi_1$    & 0&287(4) & 1&810(18) & $-$0&016(17)  & 0&10 \\
$\Phi_2$    & 0&515(8) & 0&341(2)  &    0&0004(17) & 0&11\\
$\Phi_3$    & 0&567(8) & 1&031(8)  &    0&003(5)   & 0&11\\
$\Phi_{W1}$ & 0&449(14)& 0&881(52) &    0&0004(25) & 0&078 \\
\hline\hline
$am=0.07$ & \multicolumn{2}{c}{$aM_i$} & \multicolumn{2}{c}{$A_i$}
          & \multicolumn{2}{c}{$A_0$} 
          & \multicolumn{2}{c||}{$\sqrt{(A_0+2\sigma)/A_i}\leq $} \\ \hline
$\Phi_1$    & 0&288(3) & 1&805(12) & $-$0&0095(97) & 0&074\\
$\Phi_2$    & 0&515(7) & 0&3411(3) &    0&0003(12) & 0&089\\
$\Phi_3$    & 0&566(7) & 1&031(8)  &    0&0020(34) & 0&092\\
$\Phi_{W1}$ & 0&448(12)& 0&880(50) &    0&0002(15) & 0&06\\ \hline\hline
\end{tabular}
\end{center}
\caption[]{\label{2exfit} \it
Results of double exponential fits with a fixed assumed light mass to
the lowest states in the confinement phase ($\beta_G=12$,
$\beta_H=0.3411$, $L^2\cdot T=40^3$). }
\end{table}

The amplitude $A_0$ is consistent with zero in all cases. Adding two
standard deviations to $A_0$ we get a bound at 90\% CL for the ratios
of the amplitudes.  The square root of this ratio, which is given in
the last column of the table, may serve as an estimate of the maximal
matrix element that a lower state has with the corresponding eigenstate
of our basis.
This suggests that it is rather unlikely that significantly lighter
states have been missed.

A potential source of systematic errors in the reported values of
ground state masses is the residual contamination of the correlation
function by higher excitations. Of course, the blocking procedure in
conjunction with our variational technique is designed to optimise the
projection onto the ground state. In the vector channel, however, the
plateaux set in at larger values of~$t$, thus showing that the ground
state does not dominate the correlation function at very early
timeslices. 

In order to quantify this systematic error, we performed a double
exponential fit similarly to eq.\,(\ref{2ex}). Here, however, the
mass~$am$ was fixed to the mass estimate for the first excited state
in either the scalar or the vector channel.

Extending the fitting interval to earlier timeslices, we found that
the double exponential fit does not change at all the mass of the
lowest $0^{++}$ state, thus confirming that a nearly perfect
projection has been achieved. In the vector channel, the double
exponential fit gave slightly lower results for the mass of the ground
state. We found that the mass decreased by about 5\%, but that none of
the mass estimates using a double exponential fit were incompatible
with the result using a single exponential.

We conclude that practically all contamination from higher states has
been eli-minated in the scalar channel, while higher excitations might
lead to an uncertainty of about 5\% in the mass of the vector channel.

\subsection{The continuum limit}
\label{sec:cont}

Our next task is to extrapolate the lattice spectrum 
to the continuum by taking the limit $\beta_G \rightarrow
\infty$. The continuum limit is performed only for the lowest states
where our results are accurate enough. We have taken our results at
all three $\be_G$-values at the largest respective lattice sizes,
which, as Figure\,\ref{vol} shows, have reached, or are close to, the
infinite volume limit. 

\begin{figure}[hb]
\begin{center}
\leavevmode
\epsfysize=250pt
\epsfbox[20 30 620 730]{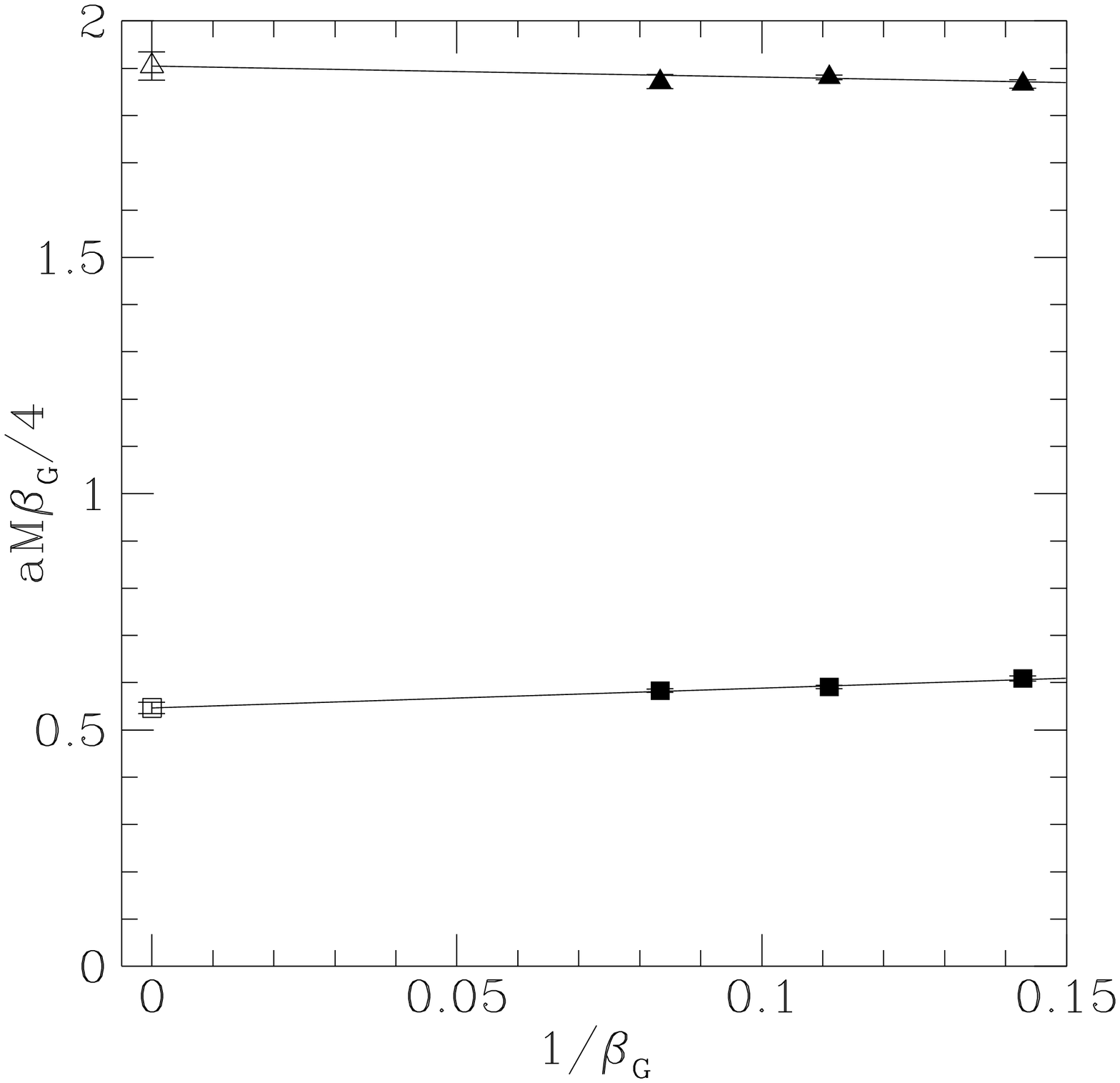}
\leavevmode
\epsfysize=250pt
\epsfbox[20 30 620 730]{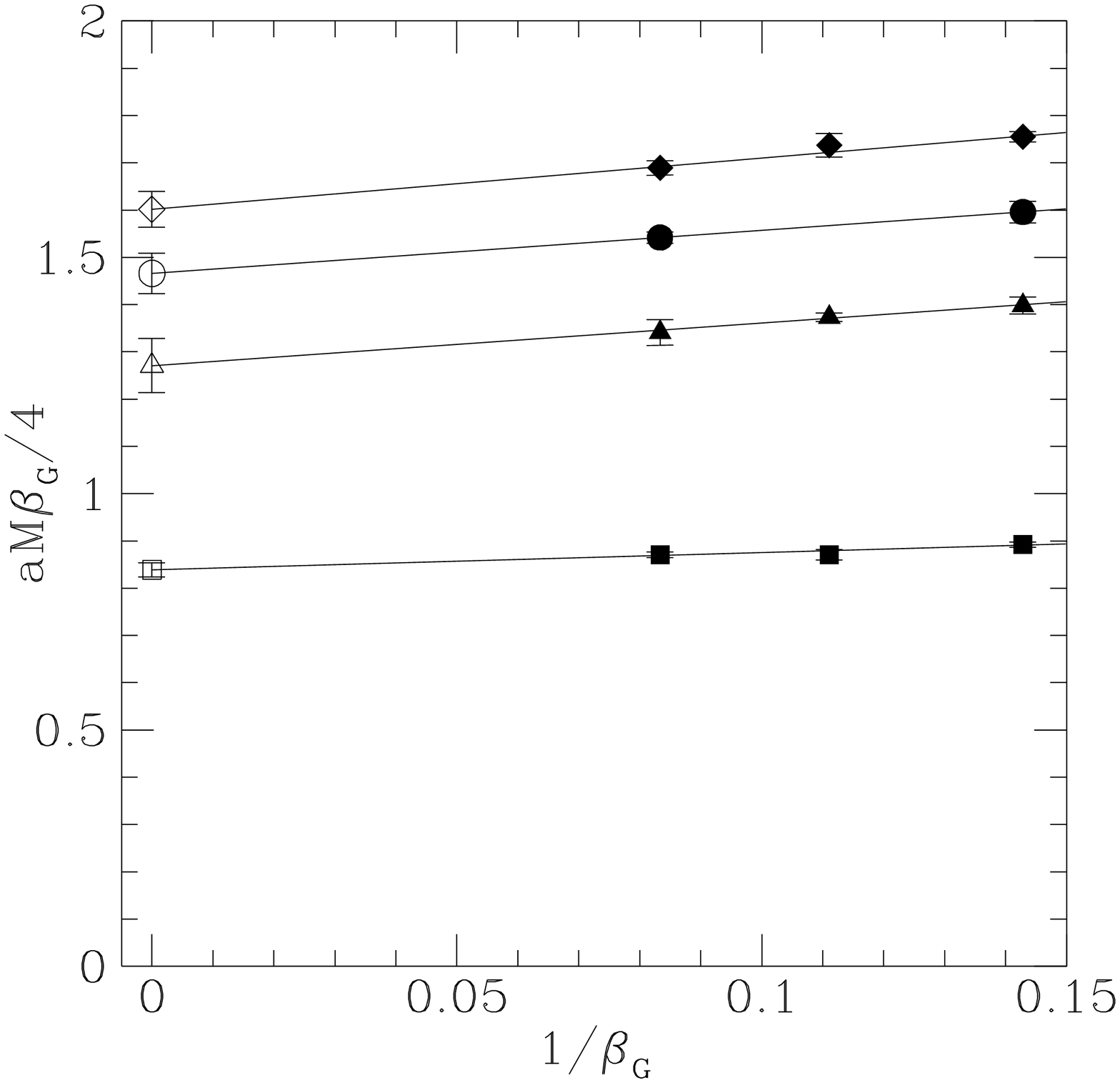}
\vspace{-1.6cm}
\end{center}
\caption[]{\it \label{contb_s}
 Continuum limit in the Higgs (left) and confinement (right) regions.
 Squares, circles and diamonds represent the three lowest
  $0^{++}$ states, whereas triangles denote the lowest $1^{--}$ state.
  Open symbols indicate the data extrapolated to $1/\beta_G=0$.}
\end{figure}

In the confinement phase, the dimensionless combinations $aM\,\be_G/4 =
M/g_3^2$ were extrapolated linearly in $1/\be_G$ for the three lowest
$0^{++}$ and the lowest $1^{--}$ states. In the Higgs phase, 
only the lowest scalar and vector states were extrapolated, since higher
excited states could not be clearly identified at all three 
$\be_G$-values. In Figure\,\ref{contb_s} the data at the three $\be_G$-values
are shown together with the extrapolated results.
In addition, we extrapolated the dimensionless ratio $aM_{H1}/aM_W$
linearly in $1/\be_G$ in both phases. Table\,\ref{tab:contb_s} shows a
summary of the continuum values of the individual masses and the
mass ratio for both phases.

\begin{table}[ht]
\begin{center}
\begin{tabular}{||l||l||r@{.}l|r@{.}l|r@{.}l|r@{.}l|r@{.}l||}
\hline
\hline
$\la_3/g_3^2=0.0239$&
  & \multicolumn{2}{c|}{$M_{H1}/g_3^2$}
  & \multicolumn{2}{c|}{$M_{H2}/g_3^2$}
  & \multicolumn{2}{c|}{$M_{H3}/g_3^2$} 
  & \multicolumn{2}{c|}{$M_{W}/g_3^2$}
  & \multicolumn{2}{c||}{$M_{H1}/M_W$} \\ \hline\hline
Higgs, & mass & 0&547(12) & \multicolumn{2}{c|}{--} 
& \multicolumn{2}{c|}{--}
      & 1&91(3) & 0&282(8)\\
$\mu_3^2/g_3^4=-0.020$ & $\chi^2/\rm dof$ & 0&80 & \multicolumn{2}{c|}{--} 
& \multicolumn{2}{c|}{--} 
                 & 1&25 & 1&70  \\
\hline 
Confinement,& mass & 0&839(15) & 1&47(4) & 1&60(4)  & 1&27(6) & 0&655(30)  \\
$\mu_3^2/g_3^4=0.089$ & $\chi^2/\rm dof$ & 0&74 & \multicolumn{2}{c|}{--} &
 0&42 & 0&06 & 0&58 \\
\hline
\hline
\end{tabular}
\end{center}
\caption[]{\label{tab:contb_s} \it 
Continuum values of the three lowest scalar and the lowest vector
states, as well as the ratio $M_{H1}/M_W$ in the Higgs and confinement
phases. Since the extrapolation of $M_{H2}/g_3^2$ was performed using
only the data at $\be_G=12,\,7$, we cannot quote $\chi^2/\rm dof$, the
error is a subjective estimate.}
\end{table}

\section{Summary and Conclusions}
\label{sec:conclusions}

We have presented results for the mass spectrum of the continuum SU(2) Higgs
model in three dimensions at selected points in the symmetric and
broken phases of the model. In order to get reliable mass estimates,
the use of improved lattice operators turned out to be crucial. This is
of particular importance for the investigation of the possibility of
very low-lying states of the kind predicted by the analytic approach presented
in \cite{bp94}.

Using our particular blocking procedure, we were able to increase the
projection onto the ground state dramatically.  In most cases in the
scalar channel, we achieved projections of essentially 100\%, whereas
in the vector channel values for the overlap ranged between 75--95\%.
Undoubtedly, with a more refined smearing or blocking procedure,
one could improve the signal for the ground state in the $1^{--}$ channel
even further. We wish to emphasise the importance of a high
projection onto the desired state, 
since otherwise the possible 
misidentification of plateaux in the effectice masses 
is a source of large systematic errors which are difficult to quantify.
Due to our use of the blocking procedure, we observe
quantitative differences in the masses of the lightest scalar and
vector states on specific lattices in the symmetric
phase compared to ref.\,\cite{ilg95}. Furthermore, we observe strong
finite-size effects in the ground state of the $0^{++}$ channel in the
symmetric phase.

Within the framework of our calculation we find no evidence for very
small masses in the scalar and vector channels in the symmetric phase,
as predicted by \cite{bp94}. 
We wish to point out, however, that we considered
correlations of gauge-invariant composite operators, whereas the
correlators of elementary fields used in the analytic approach in
\cite{bp94} are gauge-dependent. There are indications from the numerical
work reported in \cite{karsch} that correlations of these
gauge-dependent operators indeed exhibit a signal corresponding to a
very low effective mass of the gauge boson. 
This needs to be better understood.

Our computation of masses of excited states confirms the existence of
a dense spectrum of states in the confinement phase of the model.
This appears to be consistent with the picture that bound states
constitute the particle content of the symmetric phase. A surprising
result of our calculation is the existence of states that are composed 
almost entirely 
out of gauge degrees of freedom. This ``$W$-ball" and its excitations
are almost identical in mass to their gluonic counterparts in the pure
SU(2) gauge theory. We are thus led to conclude that the pure gauge
sector in the SU(2) Higgs model approximately decouples from the
scalar degrees of freedom, a phenomenon which must be of dynamical origin.

We have shown in this paper that by using various refined
calculational tools in lattice simulations of the SU(2) Higgs model,
detailed information of the mass spectrum in the symmetric phase can
be gained. 
This is important for the development of effective theories of
the symmetric phase, which will serve to analyse the nature of the
phase transition at very large Higgs masses \cite{kaj95,kar95},
and to describe the thermodynamics of the electroweak plasma 
in the high temperature symmetric phase in the 
early universe.

\paragraph{Acknowledgements}

We are indebted to C.T.\,Sachrajda for allowing us to use part of his
allocation of computer time on the Cray Y-MP at RAL under PPARC grant
GR/J86605. 
We thank R.~Sommer for fruitful discussions. 
O.~P.~acknowledges financial support by the EU ``Human
and Capital Mobility" program.


\newpage
\begin{thebibliography}{99}
\bibitem{appel}
T. Appelquist and R. D. Pisarski, Phys. Rev. D23 (1981) 2305.

\bibitem{krs85}
V. A. Kuzmin, V. A. Rubakov and M. E. Shaposhnikov, Phys. Lett. B155\\
(1985) 36.

\bibitem{fkrs94}
K. Farakos, K. Kajantie, K. Rummukainen and M. Shaposhnikov,
Nucl. Phys. B425 (1994) 67;\\
A.~Jakov\'ac, K.~Kajantie and A.~Patk\'os, Phys.~Rev.~D49 (1994) 6810. 

\bibitem{bp94}
W. Buchm\"uller and O. Philipsen, Nucl.~Phys.~B443 (1995) 47.

\bibitem{do95}
H.-G.~Dosch, J.~Kripfganz, A.~Laser and M.~G.~Schmidt,
Phys.~Lett.~B365 (1995) 213.

\bibitem{jan95}
K.~Jansen, Talk given at the International Symposium on Lattice 
Field Theory, 11-15 July 1995, Melbourne, Australia,
hep-lat/9509018, to appear in the proceedings.

\bibitem{wet93}
M.~Reuter and C.~Wetterich, Nucl.~Phys.~B408 (1993) 91;\\
B.~Bergerhoff and C.~Wetterich, Nucl.~Phys.~B440 (1995) 171.

\bibitem{kaj93}
K.~Kajantie, K.~Rummukainen and M.~Shaposhnikov, 
Nucl.~Phys.~B407 (1993) 356.

\bibitem{ilg95}
E.~M.~Ilgenfritz, J.~Kripfganz, H.~Perlt and A.~Schiller,
Phys.~Lett.~B356 (1995) 561.

\bibitem{kaj95}
K.~Kajantie, M.~Laine, K.~Rummukainen and M.~Shaposhnikov,
CERN preprint CERN-TH/95-263 (1995).

\bibitem{mon95}
Z.~Fodor, J.~Hein, K.~Jansen, A.~Jaster and I.~Montvay,
Nucl.~Phys.~B439 (1995) 147.

\bibitem{tep87}
M.~Teper, Phys.~Lett.~B187 (1987) 345.

\bibitem{tep92}
M.~Teper, Phys.~Lett.~B289 (1992) 115.

\bibitem{lai95}
M.~Laine, Nucl.~Phys.~B451 (1995) 484.

\bibitem{frad79}
E.~Fradkin and S.~Shenker, Phys.~Rev.~D19 (1979) 3682.

\bibitem{kue84}
H.~K\"uhnelt, C.~B.~Lang and G.~Vones, Nucl.~Phys.~B230 [FS10] (1984) 31;\\
J.~Jers\'ak, C.~B.~Lang, T.~Neuhaus and G.~Vones, Phys.~Rev.~D32 (1985)
2761;\\
W.~Langguth and I.~Montvay, Phys.~Lett.~B165 (1985) 135.

\bibitem{fabhaan} 
K.\,Fabricius and O.\,Haan, Phys.~Lett. B143 (1984) 459.

\bibitem{kenpen} 
A.D.\,Kennedy and B.J.\,Pendleton, Phys.~Lett. B156 (1985) 393.

\bibitem{bunk_lat94}
B.\,Bunk, Nucl.~Phys.~B (Proc. Suppl.) 42 (1995) 566.

\bibitem{fodjan}
Z.\,Fodor and K.\,Jansen, Phys.~Lett. B331 (1994) 119.

\bibitem{albanese} 
M.\,Albanese et al., Phys.~Lett. B192 (1987) 163; Phys.~Lett.~B197
(1987) 400. 

\bibitem{QCD_smearing}
S.\,G\"usken et al., Nucl. Phys.~B
(Proc. Suppl.) 17 (1990) 362; Phys.~Lett. B227 (1989) 266;
C.\,Alexandrou, F.\,Jegerlehner, S.\,G\"usken, K.\,Schilling and
R.\,Sommer, Phys.~Lett. B256 (1991) 60;  Nucl.~Phys B414 (1994) 815;\\
UKQCD Collaboration (C.R.\,Allton et al.), Phys.~Rev. D47 (1993) 5128;\\
E.\,Eichten, G.\,Hockney and H.B.\,Thacker, Nucl. Phys. B
(Proc. Suppl.) 17 (1990) 529.

\bibitem{how89}
H.\,Wittig, Nucl.~Phys. B325 (1989) 242.

\bibitem{matrix_corr}
B.~Berg and A.~Billoire, Nucl.~Phys.~B221 (1983) 109; \\
G.C.\,Fox, R.\,Gupta, O.\,Martin and S.\,Otto, Nucl.~Phys.~
B205 (1982) 188; \\
M.\,L\"uscher and U.\,Wolff, Nucl. Phys. B339 (1990) 222; \\
A.S.\,Kronfeld, Nucl.~Phys.~B (Proc. Suppl.) 17
(1990) 313.

\bibitem{seibert} 
D.\,Seibert, Phys.~Rev.~D49 (1994) 6240.

\bibitem{tepunp}
M.~Teper, unpublished.

\bibitem{karsch}
U.M.\,Heller, F.\,Karsch and J.\,Rank, Phys.~Lett. B355 (1995) 511; \\
A.\,Patk\'os, talk given at the DESY Theory workshop,
September 26-29, (1995), Hamburg, unpublished.

\bibitem{kar95}
F.\,Karsch, T.~Neuhaus and A.\,Patk\'os, Nucl.~Phys.~B441 (1995) 629.
\end{thebibliography}
\end{document}